\documentstyle[12pt,aaspp4]{article}
\def\IZw18{I~Zw~18}
\def\m82{M82}

\def\deg{\mbox {$^{\circ}$}}
\def\msun{\mbox {${\rm ~M_\odot}$}}
\def\zsun{\mbox {${\rm ~Z_{\odot}}$}}
\def\lsun{\mbox {${~\rm L_\odot}$}}
\def\msunyr{\mbox {$~{\rm M_\odot}$~yr$^{-1}$}}
\def\angs{\mbox {~\AA}}
\def\Ha{\mbox {H$\alpha$~}}
\def\Hb{\mbox {H$\beta$~}}

\def\line{\mbox {~$\lambda$}}
\def\lines{\mbox {~$\lambda\lambda$~}}
\def\o3hb{[OIII]$\lambda5007$~/~H$\beta$~}
\def\O1ha{[OI]$\lambda6300$~/~H$\alpha$~}

\def\s2ha{[SII]$\lambda\lambda6717,31$~/~H$\alpha$~}
\def\2z2{HeII~$\lambda4686$~}
\def\z7{[NII]~$\lambda6583$ }
\def\N2{[NII]~$\lambda6583$~/~H$\alpha$~}
\def\16z2{[SII]~$\lambda\lambda6717, 6731$ }

\def\n{NGC~}
\def\asec{\ifmmode {'' }\else $''~$\fi}  
\def\amin{\ifmmode {' }\else $'~$\fi}    
\def\arcsper{\ifmmode \rlap.{'' }\else $\rlap{.}'' $\fi} 
\def\arcmper{\ifmmode \rlap.{' }\else $\rlap{.}' $\fi} 
\def\sles{\lower2pt\hbox{$\buildrel {\scriptstyle <}
   \over {\scriptstyle\sim}$}} 
\def\sgreat{\lower2pt\hbox{$\buildrel {\scriptstyle >}
    \over {\scriptstyle\sim}$}} 
\def\kms{~km~s$^{-1}$~}
\def\ergsec{~ergs~s$^{-1}$~}
\def\sb{~ergs~s$^{-1}$~cm$^{-2}$~arcsec$^{-2}$}

\def\cm3{~cm$^{-3}$}

\def\fig{{Figure}}

\def\x{{X-ray}~}

\def\et{{\rm et\thinspace al.}\ }   

\def\apj{ApJ}
\def\apjs{ApJS}

\def\aj{AJ}
\def\mn{MNRAS}
\def\nat{Nature}
\def\aa{A\&A}
\def\aasup{A\&AS}
\def\baas{BAAS}
\def\annrev{ARA\&R}

\begin{document}

\title{The Impact of Star Formation on the Interstellar Medium
in Dwarf Galaxies: 
II. The Formation of Galactic Winds }


\author{Crystal L. Martin\altaffilmark{1,2,3,4}}

\altaffiltext{1}{Hubble Fellow}

\altaffiltext{2}{Steward Observatory, University of Arizona, Tucson, AZ 85721}

\altaffiltext{3}{Currently,  Space Telescope Science Institute, 3700 San Martin
Drive, Baltimore, MD 21218}

\altaffiltext{4}{
Visiting astronomer Kitt Peak National Observatory
.}

\begin{abstract}

Images and longslit, echelle spectra of the \Ha emission from 14 dwarf 
galaxies and M82 have been used to identify expanding shells of ionized gas.  
Supershells  (radius $>~300$~pc) are found in 12 of the dwarfs.
The measured shell sizes and expansion speeds constrain the ages and power 
requirements of the bubbles. The dynamical age of the larger bubbles is 
typically about 10~Myr, and ionized shells older than 20~Myr are rare.
An energy equivalent to 100 to 10,000 supernova explosions
over this period is needed to drive the shock front that sweeps out the 
cavity. The current star formation rates are high enough to meet these  power 
requirements. Many of the shells will breakthrough the surrounding layer of HI
supersonically, but the projected expansion speeds are typically less than the 
lower limits on the escape velocity. Some of the shell material
may permanently escape from a few galaxies such as \n1569.  
Whether bound to the galaxy or
not, these outflows probably play an important role in regulating the star 
formation rate and are expected to significantly influence the chemical 
evolution of the galaxies. The shells lift gas out of the disk at rates 
comparable to,  or even greater than, the current galactic star formation 
rates. They  will only displace a substantial fraction of the interstellar gas
if their duty cycle is much longer than the rotational period of the disk.  


\end{abstract}

\section{Introduction}

The interplay between massive stars and the interstellar medium (ISM)
plays a fundamental role in the formation and evolution of galaxies.
In addition to ionizing radiation and newly synthesized elements,
massive stars deliver kinetic energy and momentum  to the surrounding gas 
through stellar winds and supernova explosions.
Shock waves driven by an ensemble of massive stars may trigger
additional star formation  and/or sweep the interstellar gas out of the
region actively forming stars (Tenorio-Tagle \& Bodenheimer 1988).
The gas flows create a turbulent pressure which helps 
support the weight of the ISM (e.g. McKee 1990) and  cavities which apparently
enhance the distance ionizing radiation propagates (Hunter \& Gallagher 1997;
Martin 1997). 

This feedback from star formation may have a particularly strong 
influence on the evolution of low mass galaxies.  Owing to their
low escape velocity, Larson (1974) suggested that the loss of supernova-heated
gas would begin earlier and carry away a larger fraction of their initial
mass.  This idea was further developed  by Dekel \& Silk (1988) who
used the supernova feedback to regulate the star formation history of the
evolving dwarfs.  
Their starburst-driven wind models were consistent with
the observed mass-metallicity and mass-radius scaling relations of dwarfs
when a halo similar to those produced in cold dark matter cosmological 
simulations was included.
Mass loss  has subsequently been proposed to explain a number of peculiarities
about dwarf galaxies such as their
abundance patterns (Marconi,Matteucci, \& Tosi 1994) and rapid evolution
at moderate redshifts (Phillipps \& Driver 1995; Babul \& Rees 1992). 
The ejection of the ISM may not be as easy as previously thought, however.
In particular, the rupture of a supershell perpendicular to
a galactic disk may vent much of the energy leaving most of the disk
intact (DeYoung \& Heckman 1994).

Observations of dwarf galaxies reveal an environment conducive to
the growth of large bubbles.  Their rotation is typically nearly
solid body, so  shells are not sheared apart; and metallicities are
generally sub-solar so cooling times are longer.  Indeed, small 
bubbles permeate the star forming regions of the Magellanic Clouds,
and a hierarchy of giant shells ($R < 300$~pc) and supergiant shells 
($R \ge 300$~pc) is plainly visible
(Davies, Elliot, \& Meaburn 1976; Meaburn 1980; Kennicutt \et 1995).
The formation of regions like 30~Doradus,
which will evolve into a supergiant shell (Chu \& Kennicutt 1994),
may be thought of as the first step in the formation of a galactic outflow.
Deep imaging of the ionized gas in other dwarfs yields a plethora
of candidate structures for supergiant shells.  Indeed, roughly
one out of every four high-surface brightness dwarfs exhibit 
at least one shell  and/or filaments (Hunter \et 1993).
It is not always obvious, however, which arcs and filaments
will show the kinematic signature of an expanding shell (Hunter \&
Gallagher 1990).  

The kinematic evidence is mounting that some shells do breakthrough
the ambient neutral gas.  In the LMC, for example, the kinematics of many
supershells are surprisingly quiescent  compared to the giant shells 
(Hunter 1994).  Some of these supershells are believed to be the inner ionized
surface of cylindrical HI holes (Meaburn 1979; Meaburn 1980; Hunter 1994),
which may have formed as a superbubble blew out perpendicular to the galactic plane.  In another Magellanic irregular galaxy, \n4449, 
the very large HI hole may be associated with a shell that expanded out of the
galactic plane (Hunter \& Gallagher 1997).  In less luminous galaxies
like the blue compact dwarf \n1705, the expansion of the shell around the 
central starburst is decidedly non-spherical (Meurer \et 1992).
The kiloparsec scale, expanding shells in amorphous dwarfs (Marlowe \et 1995)
and the faint galaxy IZw18 (Martin 1996) also seem to be elongated in  
the general direction of the HI minor axis.
At issue, however, is whether any of these disk outflows develop into freely 
flowing winds in which the gas actually escapes from the
gravitational potential of the galaxy. Only one member 
of Marlowe's sample, \n3955, was a strong wind candidate.
The most convincing arguments for actual mass ejection are
based on the detection of \x emitting gas  well
above the galactic plane of \n1569 (Heckman \et 1995).
The association of a hot bubble with the cavity formed by the
expanding network of extended \Ha filaments is reminiscent of the minor axis 
outflow  from M82 (Bland \& Tully 1988; Strickland \et 1996; Shopbell \et 1997),
although it is not yet clear whether the dynamics of these two classes
of galactic outflows are completely analogous.

A more extensive kinematic census is desired to assess
the frequency of blowout and the amount of mass loss.  This paper
presents a catalog of large-scale expanding structures in 14 nearby
dwarfs.  Although M82 does not strictly meet the sample selection
criteria, it was added to the sample to provide a common galaxy
between this study and studies of superwinds from more
luminous starbursts (Heckman, Armus, \& Miley  1990, hereafter HAM).
Galaxies were selected from a volume of radius $d \le 10$~Mpc,
right ascension $4 h \le \alpha \le 14~h $, and  declination $\delta \ge -35$\deg.  An effort was made to pick the galaxies with the most intense star
formation over a range in absolute luminosity from 
$M_B = -13.5$ to $M_B = -18.5$.  
Each radial velocity field was sampled with 
deep, high-resolution spectra of the \Ha emission.
Additional properties of the galaxies are summarized in Table~\ref{tab:sam}.

While similar scale shells are found throughout the sample, the
net impact on the host galaxy's evolution may be quite varied.
Two factors which largely determine the bubble's fate -- i.e. the
distribution of the HI and the gravitational  potential -- are not
at all uniform across the sample.   Hence, the prospects for mass
ejection  are discussed on a galaxy by galaxy basis.
The results have interesting applications for the chemical evolution
of dwarf galaxies and the regulation of their star formation rate.

This paper is organized as follows.
The observations and  data reduction  are described in \S~\ref{sec:obs}.  
Section~\ref{sec:results} describes the kinematics of the ionized gas, 
and \S~\ref{sec:dynamics} discusses the dynamics of the expanding shells.
Rotation curves are sketched in \S~\ref{sec:mdot}, and the shell expansion
speeds are compared to the escape velocity.
Section~\ref{sec:sum} summarizes the main results.


\section{Observations}
\label{sec:obs}

Narrowband CCD images of the galaxies at  6580 \angs\ and  6450 \angs\
were obtained at the Steward Observatory 2.3m telescope (Table~\ref{tab:obs}).
After standard processing to remove fixed pattern noise (Rieke 1994),  
the continuum images were subtracted
from the corresponding \line\ 6580 images to produce images of the  
\Ha + [\ion{N}{2}] line emission.
These images were flux calibrated using observations of standard stars.
Photometry of the HII regions  agrees with  other published values to better than
20\% in all cases, and most of the discrepancy is thought to be caused by
slight differences in aperture. The images are sensitive to surface  brightnesses 
$\Sigma (\Ha + [NII]) \ge 4 \times 10^{-17} $\sb\ over areas of a few square arcseconds.

The structure of the ionized gas was used to select slit positions for
measuring the gas kinematics.
The number of observations and positioning strategy varied based on the size of the
galaxy and was revised during the course of each observing run based on initial
results.    In general, most of the large-scale shells and filaments were
sampled, and the spatial extent of kinematically interesting features 
was determined by follow-up observations with the slit rotated 90\deg.
The positions of the slits are identified by number on the images 
in \fig~\ref{fig:images}.

Longslit spectra of the \Ha + [NII] emission lines were obtained using the
echelle spectrograph on the KPNO 4 m telescope with the Tek 2k $\times$ 2k CCD.
The instrumental setup,  described in Martin \& Kennicutt (1995),
provided 11 \kms resolution (FWHM of the night sky lines) and a usable
slit length of $\sim 3\farcm5$. When observing conditions were good, three 
1200 sec exposures were sufficient to remove cosmic rays and reveal the line profile 
of the faintest emission detected in the \Ha images.  The CCD frames were
reduced in the standard way (e.g.  Martin \& Kennicutt 1995), and  the wavelength calibration
is accurate to better than 0.1\angs .  The narrow emission lines from the night sky have
been left in the echellograms in \fig~\ref{fig:edata} to illustrate the 
accuracy of the distortion correction and the spectral resolution.  
The spatial resolution varies from 1\asec to 2\farcs
 
\section{The Kinematic Atlas and Supershell Catalog}
\label{sec:results}

Figures~\ref{fig:edata}a-l present a kinematic atlas of the \Ha echellograms, 
grouped by galaxy and arranged in a spatial sequence.
Even a  quick inspection immediately reveals a wealth of structure 
in the velocity field.   Finding and cataloging the
large-scale kinematic structures in the warm, ionized gas is, however,
a several step process.  I searched for ``Doppler ellipses'' on
individual echellograms, then located their positions on the \Ha images.  
In almost every case, the kinematically selected regions correspond to
shells or filaments in the image. (Note that the converse of that statement
does not always hold.) Hence, a two-dimensional picture of the large 
scale line-of-sight velocity field can be built up by
combining the measurements from several slit  positions and the
morphology of the \Ha image. In this section, I draw attention to 
examples of kinematic structures at individual slit positions, then
build a picture of the large-scale kinematic structures
in each galaxy.

\subsection{Doppler Ellipses}
In \fig~\ref{fig:edata}, all regions with two or more peaks
in the \Ha line profile have been marked. 
Region~D near the west end of \n4449-9 and
feature~A along slit \n3738-2 and are good examples  of the
kinematic signature of expanding shells on small angular scales.
Their elliptical shape in the echellograms is attributed to the
Doppler shift of the emission from the approaching and receding sides 
of an ionized shell (e.g. Osterbrock 1989 \fig~6.6). 
I extend the concept of a ``Doppler ellipse'' to include features
like region~A along \n3077-5 where only partial segments of the ellipse
are visible.  {\it Polar bubbles} -- like
shell~A along \n3077-5, shell~S in \IZw18-1, or shell~B in \n4449-3 -- 
are critical to the discussion in this paper.  They
have two defining characteristics:
(1) the associated starburst is located near one end of the Doppler ellipse 
rather than at its center and
(2) the blue-shifted and red-shifted sides of the line profile have 
different intensities. In the most extreme cases, M82-2 and \n1569-10,
the magnitude of the line splitting increases with projected
height above the plane of the galactic disk, and the two velocity
components are not observed to re-converge to a single component.

The expansion velocity and diameter of each shell
in \fig~\ref{fig:edata} were estimated from the  maximum amount of
line splitting measured along the Doppler ellipse and
the length of the region of split lines respectively.  When only part
of a Doppler ellipse was plainly visible, 
an ellipse was fit to the two-dimensional line profile of the region
marked in \fig~\ref{fig:edata}, and   the expansion velocity  and
shell diameter  measured from the lengths of its axes.

\subsection{Reconstructing the Global Gas Kinematics: 
Example NGC~3077}

Figures~\ref{fig:images}a-l illustrates the location of the supershells
which have been identified and catalogued in Table~\ref{tab:bubbles}.
The solid ellipses  show the regions where the Doppler ellipses
were detected in the echellograms. Their ellipticity is indicative of the 
dynamical age of a shell.  The dotted lines outline the 
entire complex of warm-ionized gas that I associate with a single
expanding bubble.


For example on the \Ha image of \n3077 in \fig~\ref{fig:images}a,
ellipses are drawn at the location along slit position  \n3077-5 
where shells A and B are detected kinematically.
The spatial axis of the ellipse is indicated by a solid line,
and the velocity axis is scaled such that 0\farcs3 is 0.1\angs\
of Doppler shift.
A bright loop of emission extends south-eastward from the starburst,
but shell~A extends  out to a fainter loop
46\asec\ ($\sim 800$~pc)  beyond the starburst.

The perimeter of this fainter loop defines complex~A
in \fig~\ref{fig:images}.
Complex B, just west of the galaxy, is also detected kinematically 
along positions \n3077-2 and \n3077-3. Kinematic detections of the same complex
along several position angles reduce the need to resort to morphology to
constrain the area of the bubble, so the slits with kinematic detections 
are listed in col.~2 of the supershell catalog (Table~\ref{tab:bubbles}).

The filaments in complexes G and J are clearly the limb brightened edges of the
polar bubbles seen on each end of \n3077-2.
The region of  line-splitting labeled
shell~J extends  34\arcsec (590 pc) south of the starburst
and a fainter loop protrudes further to 54\arcsec (940 pc)
in the image.
These bubbles, like A and B, have apparently pushed their way out of the star 
forming region in the direction of the least resistance from the ambient gas.
In the catalog, the velocity in col.~3 is 1/2 the maximum line splitting
measured across the face of the bubble.  The projected {\it height} of each
bubble above  the starburst region is given in cols.~4 and~5.  Column~6 lists
the average \Ha + [NII] surface brightness within the complex.

In contrast, the ``radius'' listed for shell D, which is  
coincident with the bright ring of emission just north of the starburst,
is 1/2 its diameter.  The circular shape of the shell,
its higher velocity, and the presence of interior continuum emission are
consistent with a younger shell that is still roughly spherical in shape. 
Such geometrical assumptions introduce some  subjective judgement
in the cataloged radius at the level of a factor of two.
No attempt was made to correct $R$ and $v$ in Table~\ref{tab:bubbles} to a
true height and expansion velocity based on the inclination of the 
polar bubbles to our line-of-sight.


The inclination of the polar bubbles can be constrained, however,
when the gas dynamics are fully modeled on a galaxy by galaxy basis.  
Any density gradient in the shell along the polar axis leads to an 
intensity difference between the velocity components if the polar axis
is inclined to our line-of-sight
(e.g. HAM 1990; Martin 1996). 
The shape and density structure of the polar shell determine the relative
intensities and velocity offsets of the blue-shifted and red-shifted 
components of the line.
Applied to \n3077, the relative faintness of the red-shifted component
across bubbles G and J suggest these lobes are tipped into the plane of the
sky.  The polar axis of bubble A is likely in the plane of the sky.

%

\subsection{Detections of Large-Scale, Expanding Structures
in Individual Galaxies}

\subsubsection{\n4214}
Several bubbles are found to be associated with the
young  star forming region in \n4214 
(Sargent \& Filippenko 1991; Leitherer \et 1996).
About one-half the ultraviolet light from the starburst comes from
the 4 -- 5~Myr old cluster \n4214 \#1 which lies within the southern edge
of a circular ring in the \Ha emission (Leitherer \et 1996).
This ring has a diameter of 8\farcs3 (145~pc) and is identified as
complex~A  in \fig~\ref{fig:images}.
Echellogram \n4214-1 shows the \Ha emission across it splits into
two velocity components near the location of the peak continuum emission.
Hence, the ring is probably the projection of the shell surrounding an 
expanding cavity.  The shell must not be spherical, however, since
the diameter of the Doppler ellipse is over twice that of the ring in the 
image.
The line-splitting along the edge of shell~A at positions
\n4214-3 and \n4214-2 confirm that the area covered by the bubble is larger
than shell~A itself.  The obvious explanation is that 
our line-of-sight is parallel to the polar axis of a wasp-waisted bipolar
bubble, and the shell is brightest where its expansion has been restricted by 
the higher ambient density in the galactic plane.

Two polar bubbles are detected along \n4214-1 and seem to be
breaking out of region~A.  Bubble~B is surrounded by some gaseous
filaments in \fig~\ref{fig:images}, although only diffuse emission
is seen in the region of complex~C.  Some faint filaments are found
within complex~F, although the line-splitting is confined to a few patches
with a faint component redshiftd to  higher velocity.
These wisps  reach velocities $\sim 100$\kms and
are unresolved in the spatial dimension; their  
physical interpretation is less certain than that of the well-defined
Doppler ellipses.

By observing the overall tilt of each \n4214 echellogram, notice that
the rotation of the ionized gas contributes less to the
width of the integrated \Ha line profile than these shells and wisps.
The largest gradient in the central velocity of the line profile is
38\kms across \n4214-4.  The HII complex one arcminute west of the
central starburst has a lower velocity than the eastern side of the galaxy.  
Across \n4214-1, \n4214-2, and \n4214-3, 
the ionized gas at the northeast end of the slit is moving 15 -- 26\kms
faster than that at the southwest end, so the rotation axis of the ionized 
gas is closer to the major axis of the galaxy  than the minor axis.  The HI 
rotation axis and major axis are also oriented at 
PA $\approx -20\deg$ (McIntyre 1996).

\subsubsection{\n4861}
Figure~\ref{fig:images}c illustrates the
bipolar outflow discovered in \n4861.   Shells~A and C
detected along \n4861-4 coincide with webs of gaseous 
filaments extending about 1 kpc westward and eastward, respectively,
from the giant HII complex in \n4861.
The reversal in the shape of the line profile between shell~A
and shell~C -- i.e. the intensity of the blueshifted component
is higher across shell~A while the redshifted component is more
prominent across shell~C -- is consistent with the polar axes
of these two bubbles being tipped in opposite directions from
our sightline.    This outflow axis is not aligned with the
minor axis of the continuum isophotes which extend north-eastward
in the direction of the smaller HII regions visible in the image.

The kinematics across the two bright loops to the north of the HII complex 
are less impressive.
The line profile across the eastern loop shows a faint red wing, 
labeled complex~B; 
and shell~A seems to partially overlap the line-of-sight through
the western loop.

\subsubsection{M82}

The nuclear starburst in M82 drives a bipolar outflow of
extraordinary scale along the galaxy's minor axis
(Bland \& Tully 1988;  HAM 1990).
In \fig~\ref{fig:images}, the filaments extend to a projected
height of 240\asec\ (4.2~kpc) above the disk.
Residuals from the continuum light of the stellar disk 
are visible  at a PA $\sim 68\deg$ in the image, and 
dense gas in this disk  apparently confines this superwind in the galactic plane.
Echellogram \m82-2, along the polar axis of the outflow,
is clearly double-peaked on both sides of the nucleus.
The maximum separation of these components is 313\kms\ and 270\kms\,
respectively, on the northern and southern sides of the slit -- 
in good agreement with the measurements of HAM.  These higher
resolution data  show more variation in the line profile
and flux-weighted central velocity with radius than those discussed by HAM. 
However, the maximum intensity still shifts
from the redshifted component (south) 
to the blueshifted component (north) as expected from tipping the
polar axis $\sim 35 \deg$ away from our sightline (cf. HAM, Fig. 19).

Since the two components of the line profile do not converge to  a common 
velocity  along \m82-2, I differentiate this structure from a closed Doppler 
ellipse in \fig~\ref{fig:images}. The spatial axis begins where the line profile splits into
two components and continues as far as  one side of the cavity is detected.  A flag 
is drawn at the last position where both components of the double-peaked profile are
detected,  and its length represents one-half the magnitude of their velocity separation.
The filaments south of M82 do appear to  converge about 190\asec\
south of the starburst in \fig~\ref{fig:images}, so the bubble  may  be capped at the end.

The far side of the outflow was observed at a second slit position oriented 
perpendicular to the polar axis of the wind.  Echellogram \m82-4 reveals
the width of the expanding cavity and \fig~\ref{fig:images} shows its close
correlation with the morphology of the extended filaments.  The faint substructure
interior to the big Doppler ellipse appears to form at least three and possibly four
smaller ellipses, so the wind is apparently composed of  several adjacent cells.
These interior walls give the superwind a cellular structure, and I suspect they
may be formed by smaller bubbles merging together to form the outflow.

\subsubsection{\n3738}

Three Doppler ellipses were identified along \n3738-2.  
Of these, shell~A has the highest surface brightness and largest
velocity, 37\kms,  along our line-of-sight. Complex~A was associated 
with the bright ring of \Ha emission in \fig~\ref{fig:images} because
an even larger region of line-splitting is detected kinematically along \n3738-3.
The filaments that Hunter \& Gallagher (1990) labeled 1 and 2 in their 
echellogram are probably  the signature of this expanding complex.
The radius of bubble~A in Table~\ref{tab:bubbles} 
is one-half the diameter of this shell, or $R = 8\farcs9$ (170~pc). 
The Doppler ellipse labeled shell~B coincides
exactly with a faint \Ha ring of diameter 6\farcs8 (131~pc).
The radius of bubble~C, 14\farcs9 (288~pc),  is  taken as the projected distance
from the brightest HII region to 
the faint arc just beyond the end of the Doppler ellipse.

\subsubsection{\n2363}

\n2363 is a giant HII region on one end of the dwarf galaxy \n2366.
Roy \et (1991) have mapped the line profile of its [OIII]~$\lambda 5007$
emission using a scanning Fabry-Perot interferometer.
Echellogram \n2363-4 intersects their expanding shell and 
associated chimney.  Surprisingly this echellogram shows no line-splitting.
The mean velocity of the \Ha line does, however, increase by 43\kms\ from south
to north; and this velocity shear is similar to that
measured in  [OIII]~$\lambda 5007$ by Roy \et  
The wings of the line profile  broaden to $\sim 500$\kms~FWZI 
at positions a few arcseconds north and south of the peak \Ha intensity.
These two sources are only marginally resolved.  My spectra are not very
sensitive to extremely broad lines (e.g. Roy \et 1992), since the
dispersion is high and the spectral coverage is only $\sim 4000$\kms.

Along \n2363-5 line-splitting is plainly visible along the base of 
the Roy \et chimney -- shell~B in \fig~\ref{fig:edata}.
In \fig~\ref{fig:images}g, the bright pair of filamentary loops north-northwest 
of the HII region define  the boundaries of the associated bubble,
complex~B. Echellograms \n2363-6 and \n2363-5 bisect these loops.  Although
they show line splitting of magnitude
 ${{1}\over{2}} \Delta v \approx 40$\kms across the eastern half of 
this complex, the Doppler ellipses extend much further eastward across
the fainter arcs labeled bubble~A.  The limb of shell~A is 34\asec\ (590~pc)
from the eastern HII region and also shows a double-peaked
profile along \n2363-5 where ${{1}\over{2}} \Delta v \approx 16$\kms.

\subsubsection{\n2537}

The Doppler ellipse along \n2537-1 coincides with the 
bright U-shaped HII region on the western side of \n2537.
Close inspection of the \Ha image reveals a faint nebulosity
extending 22\asec\ (800~pc) westward. A patchy Doppler ellipse 
is seen over this entire region in \n2537-3, so the faint emission comes
primarily from an expanding shell.

\subsubsection{\n1800}

Deep \Ha images show a spectacular network of filaments along the
minor axis of \n1800 (Hunter, Hawley, \& Gallagher 1993;
Hunter, van Woerden, \& Gallagher 1994). In \fig~\ref{fig:images}h, they
extend up to 34\asec (1350~pc) north of the nucleus (the peak red continuum 
emission).
Echellogram \n1800-2, aligned parallel to the  galaxy's major axis, traverses
the bright ends of these filaments where they meet the galaxy and reveals
a single Doppler ellipse 20\asec in diameter.  The expansion velocity at the
base of the northern polar bubble is 27\kms. 
The magnitude of the line splitting reaches  50\kms and
43\kms in the north and south lobes respectively (Marlowe \et 1995)


\subsubsection{Sextans~A}

As the only member of the Local Group in the sample, Sextans~A
provides an opportunity to examine the kinematics on a finer scale.
Echellogram SexA-1 bisects the  central 22\asec\ (140 pc) $\times\ $ 35\asec\
(222~pc) ring of ionized gas  along its long axis (cf. \fig~\ref{fig:images}i).  
Three regions of line-splitting are identified.  In the northern part of the ring,
Doppler ellipse~A extends +69\kms toward  longer wavelengths and -48\kms
toward lower velocities.  Hunter \& Gallagher (1992) have previously
identified the red half of this shell.  Doppler ellipse~B partly overlaps
the blue-shifted component of ellipse~A  but extends to -84\kms, while its
redshifted component, visible in a low contrast image stretch, extends to 
only +40\kms.  It is not absolutely clear whether these two ellipses
are produced by the same expanding structure.  They are catalogued as one
expanding complex, $v = 1/2 \bigtriangleup v \approx 60$\kms, primarily
because of their association inside the the ionized ring in \fig~\ref{fig:images}i.  The broad, FWHM~$\approx 2$\angs, wisps of emission in
Region~C are separated from the main component of the line profile by
+90 and -110 \kms.

\subsubsection{I~Zw~18}
A bubble with radius $r = 970$~pc and projected expansion velocity 
$v = 34$\kms was found south-southwest of the northwest HII region.
A kiloparsec-scale loop of \Ha emission also extends north-northeastward
from it, although the kinematic evidence for expansion is less secure.
Notice how similar the velocity structure in I~Zw~18-1 is to 
\n1569-11, \n1569-10, and even to M82-2. The main difference is  that the
magnitude of the line splitting is about five times larger in M82.
See Martin (1996) for a discussion of the power requirements of these shells 
and the associated production of metals.

\subsubsection{Non-detections}

Expanding shells were found in all but
3 of the 15 galaxies examined spectroscopically.
Although it is difficult to quantify the efficiency of the
shell-finding strategy, the high incidence of ``shell-like''
structures in the images with the kinematically detected
shells leads me to believe that few of the large shells
were missed.  In particular, II~Zw~40 and VII~Zw~403 are
unlikely to contain any supershells.
The census is, however,  incomplete in several ways.
First, structures smaller than the angular  resolution element
were obviously not detected.  Second, 
faint shells  along the sightline to high surface brightness HII regions 
could be lost in the wings of the emission from the HII region.
Third, low intensity emission spread over more than $\sim 2000$\kms
may masquerade as continuum and/or be lost in the detector noise.
Finally, although the spatial sampling appears to be sufficient to
find the largest shells, some medium size shells must be missed due
to the incomplete spatial coverage.

The null detection in \n5253 is attributed to this latter effect.
Marlowe \et (1995) did find
an expanding shell in \n5253 which my slit positions simply
miss.  They measure an expansion velocity of 35\kms east of the
starburst along PA = 60\deg; and  their \fig~3 clearly shows 
line splitting out to $R = 43\farcs8$ (870~pc) from the central starburst.  
These parameters and my measurement of the bubble's surface brightness
are included in Table~\ref{tab:bubbles}. 
I have also overlaid the slit positions of my three echellograms and
those  of Marlowe \et on \fig~\ref{fig:images}j.
(Note that my \n5253-3 and the PA = 120\deg\
observation of Marlowe \et coincide.)  While all five of the longslit
observations may have missed the region of maximal expansion on the
eastern side, the lack of any Doppler ellipse detection on the western
side is very surprising given the spatial sampling.  
The western filaments are apparently quite quiescent.
The average \Ha + [NII]
surface brightness in the region outlined in \fig~\ref{fig:images}
is $1.89 \times 10^{-16}$\sb.

\subsection{Blowout in \n1569?}



The spectacular emission-line filaments emanating to the north and south of
\n1569 were discovered
on early photographic plates (Hodge 1971; Zwicky 1971) and shown to have
residual velocities of -60\kms and +60\kms respectively
(de Vaucouleurs, de Vaucouleurs, \& Pence 1974). 
Kiloparsec scale spurs of diffuse \x emission are associated with this system
of filaments,  and approximately one-half the keV \x emission
comes from a halo 3\farcm8 by 2\farcm2 in size (Heckman \et 1995).
The nonthermal radio emission associated with the halo has a high-frequency
cutoff attributed to synchrotron radiation losses following the peak
starburst activity (Israel \& deBruyn 1988).  Heckman \et (1995) examined
the emission-line profiles along two position angles
with a resolution of about 57\kms at \Ha, and argued that the outflow 
was energetic enough to eject most of the interstellar medium.
To study the dynamics of this outflow in more detail, I obtained
better spatial coverage of the velocity field at high resolution.
The echellograms presented here provide new measurements of:
(1) the covering angle of the outflow, 
(2) the changes in the velocity field with height above the
disk,
(3) the cellular structure in the outflows, and
(4) the velocity field near super star cluster~A.

In \fig~\ref{fig:images}k,
echellograms \n1569-10 and \n1569-11 cover the outflow in an
``X'' pattern centered about 22\asec N-NE of super star cluster A.
The continuum source in the center of \n1569-11  is super star cluster A.  Just
to either side of it, the line profile has a broad blue wing extending
$\sim 160$\kms to lower velocity. A  Doppler ellipse is seen on the redward 
side.  
A shell associated with super star cluster A has apparently ruptured at least on
the approaching side. The outflow along the chimney walls is coincident with
the ring of \Ha emission inside
the  central HI hole  (Israel \& van Driel 1990).
Several \x point sources are also found nearby (Heckman \et 1995).

Beyond the starburst region,
these two echellograms show very similar amounts of line-splitting over
arcminute scales.    The kinematic signatures of shell~B along \n1569-11 and
shell~D along \n1569-10 are almost identical as are those of shell~A and
shell~G to the southwest and southeast respectively.
The shape of the double-peaked line profile is reversed in the two 
hemispheres however.  To the south the redshifted component is stronger,
but it is weaker than the blueshifted component to the north.  
Tilting the southern and northern lobes toward and away from our 
sightline, respectively, will produce such a difference if the shell
emissivity declines with scale height.

Four slits oriented perpendicular to the polar axis sample
the velocity field in the extended lobes.
Three Doppler ellipses were independently identified on
each of the three slits traversing the southern lobe.
Plotted on the image, the Doppler ellipses stack up inside three
shells, which can be followed from slit~7 to slit~4 to slit~13 at a projected
height of $640$~pc.
Three Doppler ellipses were also found along the slit traversing the northern 
outflow, so three superbubbles may have broken through the HI disk.  
The structure of the outflows is clearly more complex than a web of warm 
filaments wrapped around a single hot cavity. 
(In \n1569-7, notice also the double shell/chimney structure across the bright 
arm of shell~A.)

The deeper exposure at \n1569-13 was obtained to
examine the magnitude of the line-splitting as a 
function of height above the disk.
The largest radial velocities are found in complex~A where $v$, defined as 
one-half the maximum separation between the lowest and
highest velocity components, increases from 70\kms 140~pc south of the major
axis to 99\kms at $h=460$~pc.
Slit~13 crosses complex~A where the two most
prominent filaments are converging, but the faint emission between the arms
splits into two velocity components separated by $v= 140$\kms.
While the curvature of a shell could create a gradient in the projected 
expansion velocity of this magnitude, it is clear that
the shell is not strongly decelerated as it pushes through the galactic halo.

The emerging picture has the northern outflow tipped away from our sightline
and bubble~A on the near side of the inclined disk, but  
the line profiles along complex~G and~D complicate this simple interpretation.
Across much of complex~G (e.g. \n1569-4 and \n1569-13),
the redshifted component of the double-peaked line is consistently weaker 
than the blueshifted component, and they are comparable across complex~F. 
The projected velocity difference increases from
48\kms (\n1569-4) to 77\kms (\n1569-13) across complex~G but is constant
at $\sim 64$\kms across complex~F.   
One interpretation is that the polar axes
of bubbles~G and~F are not parallel to bubble~A.


Doppler ellipses from complexes~A and~B were detected by Heckman \et (1995)
along their
PA=70\deg\ and PA=160\deg\ spectra respectively. Some line-splitting
is also seen where these slits cross bubbles G, F, and E.  Our estimated
velocities are generally similar.  However, the echellograms show
a maximum velocity separation of only $2v = 160$\kms
in the region of complex~B (along \n1569-11) where H95 report 260\kms.

These polar outflows cover a large fraction of the projected area of
the galactic halo.  The Doppler ellipses found
along the slits oriented perpendicular to the polar axis 
contain almost all the line-emission in these regions.
The angle subtended by these expanding shells 
can be measured from their location in \fig~\ref{fig:images}.
From the vantage point of cluster~A the northern and southern  bubbles both 
subtend about  $\sim 140 $\deg.

The radii of these shells in Table~\ref{tab:bubbles} are the projected
distance of the kinematically detected shell from cluster~A.
The  filaments extend beyond this region and converge toward a curved arc
86\asec\ (920~pc) to the south.  The south-southeast spur of \x
emission appears to be contained within this structure (Heckman \et 1995).
Since the velocity separation of the two components in the line profile is still 
increasing when the fainter side fades below our detection limit, 
the shell may have partially ruptured near the top of the bubble.
In Figure~8 of Hunter, Hawley, \& Gallagher (1993), 
a fainter arc is visible at $R \approx 116$\asec (1.24~kpc); and
diffuse emission extends  beyond it to $R \approx 190$\asec (2.0~kpc).
The relation of these structures as well as
the isolated blob of \x emission nearby (Heckman \et 1995)
to the expanding bubbles is not yet clear.


\subsection{\n4449}

Excluding M82, \n4449 has a higher B-band luminosity than any member of the sample, 
and its HI mass is actually twice that of M82.  While it is awkward to call it
a dwarf galaxy,  I include it here because it provides a particularly dramatic
example of the disturbances caused by a burst of star formation in a young
disk galaxy.  The galaxy shown in  \fig~\ref{fig:images}h has a diameter 
of 5\farcm6 (5.86 kpc) at the 25 B-mag per square arcsecond isophote which
is a mere speck in the HI disk which is $\sim 13$ times larger
(Bajaja, Huchtmeier, \& Klein 1994).  Much of the ongoing star formation is taking
place in a central bar-like region running southwest to northeast, although the
recent increase in the star formation rate north of the bar (e.g. Hill \et 1994)
gives the \Ha emission a ``T'' shape.  The entire galaxy is forming stars at
the modest rate of about 0.07\msunyr, but the star formation rate in 
complex~G alone is comparable to that in all of I~Zw~18 or \n3738.

Although many filaments and partial shells appear in the \Ha image in
\fig~\ref{fig:images},
the location of the larger kinematic complexes is not immediately obvious.
Echellograms~1 and~2 were aligned parallel to the bar and offset to its NW side.
The Doppler ellipses found on the southern half of these slits are 
associated with shell~A, and the peaks in the line profile are separated by as 
much as  160\kms.  The spatial extent of the line splitting increases from 
about 30\asec at a projected ``height'' above the bar of 10\asec (170~pc) to 
approximately 45\asec at a height of 24\asec (420~pc).  The double-peaked line
profile is particularly prominent where these slits intersect a bright filament
previously noticed by Hunter \& Gallagher (1990).   The spatial axis of 
\n4449-3 follows this filament and shows a Doppler ellipse extending from the bar 
to a projected height of $\sim 40\asec$.  The projected expansion velocity, 35\kms,
is consistent with the velocity difference measured by Hunter \& Gallagher (1990). 
The intensity of the blueshifted  component is stronger which suggests a polar
axis tipped away from our line-of-sight.  
The geometry may be more complicated than a simple polar outflow, however, as
\n4449-3 also shows multiple high velocity {\em wisps} across the bar. The
wisps extend up to +175\kms from the intensity-weighted average velocity.
Their envelope has an elliptical shape (labeled feature B),  but the 
wisps do not form a clean Doppler ellipse.  Their origin is unclear, so they
have not been associated with an expanding complex in \fig~\ref{fig:images}l
or Table~\ref{tab:bubbles}. 

Southeast of the bar along echellogram \n4449-7,  Doppler ellipse~I is visible 
across the \Ha cavity.  The measured expansion velocity, 50\kms, is consistent
with the magnitude of the line-splitting Hunter \& Gallagher measured across
their filament~2.
Farther south, ellipse~H extends to 43\farcs9  between two bright filaments.
The kinematic signatures of complex I and H are clearly those of
expanding shells, and shell-like features are seen in the image. 
The HI column is also very low within complex~I (Hunter \& Gallagher 1997).
To the north of these bubbles,  patchy line-splitting and wisps are
detected up to Doppler ellipse~J.  It is no clear whether this ellipse
is related to complex~I or even complex~C.

The kinematic activity is not limited to the
region near the bar.  Echellograms \n4449-9  and \n4449-11 show Doppler 
ellipses at the NW end of the ``T'', hereafter complex~D.  A larger region of 
line-splitting labeled ``C'' in \fig~\ref{fig:images} shows several adjacent 
Doppler ellipses over $D \approx 25\asec$ region.  At the top of the ``T'', 
complex~G  extends from two HII complexes in the west to a faint arc on its eastern 
boundary.  Along \n4449-9, Doppler ellipse~G is prominent in the [NII] $\lambda$6458
line, but the \Ha emission is difficult to separate from the bright 
HII region emission.
Along the eastern perimeter of the galaxy, the three Doppler ellipses  
identified along \n4449-10 coincide with partial rings of emission in the \Ha 
image.

All the Doppler ellipses identified on the echellograms are plotted on the
\Ha image in \fig~\ref{fig:images}.  The associated kinematic complexes are
outlined by dotted lines, and several deserve further comment since their boundaries
are appreciably less well-defined than the other cataloged structures.
Features B and J show very disturbed gas motion but not clean 
Doppler ellipses, so they were not catalogued as expanding shells.  
Only the faint, fairly diffuse, emission from complex~G and
complex~D shows the kinematic signature of expansion, so the brighter regions
were masked out of the apertures for shell photometry.
The spatial extent of complexes A, C, and H were estimated conservatively.
Complex~A, restricted to the filaments comprising the Doppler ellipses,
corresponds very closely to a hole in the HI emission discovered by
Hunter \& Gallagher (1997).  Arc-shaped filaments extend further, about 1.2~kpc,
to the northwest but appear relatively quiescent (Hunter \& Gallagher 1997).
As suggested by Hunter \& Gallagher, these are probably the inner ionized edge
of  HI cavities.
The \x emission in both the soft (0.12 -- 0.28 keV) and hard (0.76--2.02 keV) ROSAT
bands does extended in the direction of bubble~A well beyond the conservative radius
of the Doppler ellipse, 34\farcs5 (600~pc) (Della Ceca \et 1997).
Several bubbles may be merging into a superbubble in the region of complex~C.
The \x map of Della Ceca \et (1997) also shows two
local maxima within complex~C.  The expansion associated with Doppler ellipse~H
probably extends further to the northwest to include the loop of emission
identified by Hunter \& Gallagher (1990, i.e. Loop~1).  Their Figure~2 panel~b 
shows hints of a Doppler ellipse in the faint component of the line.  
A prominent finger of very soft \x emission appears to curve 
southward from this complex (Della-Ceca \et 1997).  
Overall the picture is one of large shells  breaking out of the bar-like
region, and many smaller shells growing along the periphery
of the disk.

\subsection{Summary:  A Catalog of Superbubbles}

Figure~\ref{fig:rv} illustrates the distribution of bubbles in the radius
-- velocity plane.  Inspection of \fig~\ref{fig:rv} or Table~\ref{tab:bubbles}
draws attention to the relatively large number of shells with $R \ge 400$~pc.
The bubbles in seven amorphous dwarf galaxies
discussed by Marlowe \et (1995) have very similar properties.
This distribution of cataloged
shells is in fact bimodal with maxima at $R < 200$~pc and $R \approx 850$~pc.
Although the distribution of shell sizes in the Magellanic Clouds 
has a similar minimum near 750~pc (Meaburn 1980), 
I find no evidence for bimodality in the morphologically-identified
shell sample of Hunter, Hawley, \& Gallagher (1993). 
One expects my distribution of detected  shells to be strongly
biased towards the larger complexes since my sampling technique
is fairly complete on large scales but may miss many smaller bubbles. 
The apparent bimodality may therefore be misleading, although it might
be used to motivate a kinematic survey with two-dimensional spatial coverage.


\section{Superbubble Dynamics}
\label{sec:dynamics}



This section compares the power requirements and ages of all the large-scale
expanding structures in the catalog.  
The estimates for individual bubbles are only accurate to factors of a few
since the dynamical models ignore the complicated geometry.
Two model-independent measurements 
provide important checks on the validity of the results.
The integrated \Ha luminosity of each galaxy directly measures the
massive star formation rate, and areal \Ha photometry of the extended shells
indirectly constrains their mass.  For individual galaxies, hydrodynamical 
simulations tuned to the real, non-spherical distribution of gas and mass
should yield results  accurate to be better than a factor of two in the future.

\subsection{Mechanical Power from Massive Stars}

The interated \Ha + [NII] flux measured from narrowband images obtained at
the Steward Observatory Bok Telescope is given in Table~\ref{tab:star}.
The [\ion{N}{2}]\lines 6548,6584 /\Ha ratio and Balmer decrement were measured 
from longslit
spectra (Martin 1997), and luminosity-weighted corrections were applied to 
obtain the  integrated \Ha luminosity.  Column~5 shows the corresponding
Case~B recombination rate at  an electron temperature of $10^4$~K. 
Assuming no ionizing radiation escapes the galaxy, the recombination rate is 
equal to the ionization rate -- a measure of the number of massive stars.  For 
a stellar metallicity of 0.25\zsun\ and  stellar initial
mass function with slope $\alpha = -2.35$ from 1\msun\ to 100\msun,
a star formation rate of 1\msunyr\ produces $3.16 
\times 10^{53}$ hydrogen ionizing photons per second
(Leitherer \& Heckman 1995; LH95).    Column~6 of Table~\ref{tab:star}
lists the star formation rates derived from the global recombination
rate.  In the LH95 starburst model, the mechanical power reaches an equilibrium
value  $L_w = 1.45 L_{{\rm H}\alpha}$ after 40~Myr ($T_e = 10^4$~K).

\subsection{Shell Mass}
\label{sec:smass}

To photometrically measure the mass in a shell, one needs to know the
shell volume and the filling factor of warm ionized clouds.
The volume is well constrained by the projected area of the 
kinematically identified filaments, but the volume filling factor $\epsilon$
is not immediately known to even order of magnitude accuracy.  In this
paper, $\epsilon$ is treated as a free parameter, but constraints from
additional observations will be described in detail in a forthcoming paper.

The line flux from each expanding complex was measured within the aperture shown
in \fig~\ref{fig:images}.  These photometric apertures encompass mainly the extended
filaments, so contamination from foreground/background HII regions was not generally
a concern. \n4449 was an exception, and HII regions defined by a  surface brightness
threshold were masked out of the apertures for \n4449.  The fluxes were
corrected for foreground Galactic extinction, A(6570) = 0.08~mag~(csc b - 1),
and [NII] emission. The root mean square electron density for the complex is then 
$n_{rms}^2 \equiv F_{H\alpha} / (h \nu \xi \alpha_{H\alpha}^{eff}(T))
(4 \pi d^2 / V_P)$, where the number of H nuclei per free electron, $\xi$, is
$\sim 10/11$ since most of the He is singly ionized (Martin \& Kennicutt 1997),
and the recombination coefficient is evaluated at $T_e = 10^4$~K (Osterbrock 1989).
The mass of warm hydrogen and helium in the shell, $M = \mu_e m_H n_{rms} 
V_E \epsilon^{1/2}$, is sensitive to the fraction of the bubble volume filled by 
warm, ionized filaments.  The mass estimates in Table~\ref{tab:bubbles}
parameterize the filling factor in terms of $\epsilon = 0.1$, 
a value near the high end of the range 
measured in individual HII regions (e.g. Kennicutt 1984), and  
are intended to be upper limits.


When $\epsilon$ is better constrained, the bubble volume will be the major 
source of error in the mass estimate. Two volume estimates are used here to 
provide some insight into this systematic uncertainty in the shell mass. 
A volume, $V_P$, was derived from the area of the photometric
aperture times two-thirds the aperture width.
For some shells such as \n3077-D, this photometric aperture includes the entire
complex, but for polar shells such as \n3077-J the base of the shell
emerging from the starburst region is clearly not included in the volume
estimate.  The photometric aperture also overestimates the volume
in some cases, since it must extend somewhat beyond the shell to capture
all the flux -- e.g. \n4449-E1. 
To estimate the uncertainty in the mass, shell masses were calculated for
both the volume $V_P$ and an ellipsoidal model for the bubble volume, $V_E$.
The length of the major axis of the prolate ellipsoid
was set equal to the shell diameter -- often the height of the polar axis; 
and  the minor axis length set equal to the shell width.  
The mass differenence is expressed as a fraction of $M(V_E)$ in parentheses in
Table~\ref{tab:mdot}.

In column~7 of Table~\ref{tab:bubbles},
the kinetic energy of each complex   is
the product of a complex's ionized mass and half the square
of its projected expansion velocity.
Ignoring any additional kinetic energy in large, neutral shells
(e.g. Meaburn 1980), 
the kinetic energies in Table~\ref{tab:bubbles} are expected to 
be accurate to a factor of a few.  They
provide an empirical measure of the kinetic energy in the large-scale 
expanding complexes that is independent of the details of a dynamical
interpretation.


\subsection{Timescales and Power Requirements}
\label{sec:power}


%
%
%

The dynamics of a wind-driven shell provide
some basic insight into the ages and power requirements of the
supershells.  In the simplest model, ejecta from a concentration
of stellar winds and supernovae push the surrounding interstellar
gas away and generate a shock front.  This shocked ISM radiates
and collapses into a thin shell.  A second shock propagates inward
through the lower density wind of ejecta creating a very hot, low density
bubble.  This bubble's pressure drives the shell outward.  Neglecting
radiative losses from this cavity, the solution for the shell's
equation of motion is (Weaver, McCray, \&
Castor  1977; Shull 1993): 
\begin{equation}
R = 0.762 (L / \rho) ^{1/5} t^{3/5}
\end{equation}
or
\begin{equation}
V = 0.458 (L / \rho)^{1/5} t^{-2/5}.
\end{equation}

For a measured shell radius and velocity, these equations define
an age $\tau$ and mechanical power $L_w/n_0$, where $n_0$ specifies 
the ambient hydrogen density 
$n_0 \equiv \rho / (1.4 m_H)$.  
In Figure~\ref{fig:weaver}, dashed, diagonal lines illustrate  
isochrones for $\tau(R,V) =  1, 5, 10, 15, {\rm ~and~} 20$~Myr.
For a specified ambient density, the energy injection rate is
read off the orthogonal axis.  In other words, a shell in a 
uniform medium  evolves along a dotted curve from upper left
to lower right when the power supply is steady.
The inferred ages and power  of each kinematically-detected
complex are tabulated in Table~\ref{tab:bubbles} columns 9 and~10.

Several conclusions can be drawn from the locus of supershells in 
\fig~\ref{fig:weaver}. Many of the larger shells have ages $\sim 10$~Myr.
Since the shells are predominately photoionized by massive stars with lifetimes
of only a few million years (Martin 1997), the formation of massive
stars has continued for at least several generations in the starburst region.
However, no {\em ionized} shells with ages greater than $\sim 20$~Myr were
identified. The energy injection rates range from 
$L_{in} / n_0  = 10^{38}$~cm$^3$\ergsec to $ \sgreat\ 10^{42}$~cm$^3$\ergsec.
Winds from normal O stars supply $L_{in} \sim few \times 10^{36}$\ergsec 
(e.g. Chiosi \& Maeder 1986), and the contribution from an average supernova 
(i.e. $10^{51}$~ergs / 10~Myr) is similar. Hence, 
many thousands of stars contribute to the formation of the larger bubbles.
A typical supershell (800~pc, 50\kms) requires the combined
energy of $\sim 10^3$ supernovae.  
The feedback is apparently hierarchical on global scales!  


The polar, rather than spherical, geometry of many of the larger shells
affects the accuracy of these power and age estimates.
No corrections for inclination have been applied, so the measured radius and 
velocity are systematically lower than the height and maximum expansion 
velocity of a polar shell. Over such large scales,
gradients are also expected in the ambient density.
Considering these deviations from the basic model,
the parameters derived from the idealized, wind-blown bubble model
are probably instructive at the level of a factor of a few.

\fig~\ref{fig:ke} illustrates one method I have used to monitor the
growth of systematic errors.
In the wind-driven bubble model, the kinetic energy of a swept-up shell is
$KE = 1.44 \times 10^{52} n_0 R_{100}^3 v_{100}^2$~ergs,
where $R_{100}$ and $v_{100}$ are the shell radius and velocity in
units of 100~pc and 100\kms respectively. In \fig~\ref{fig:ke},
this estimate is compared to the mass of ionized gas 
estimated photometrically from the \Ha luminosity of the complex.
The requirement of consistency constrains the two parameters $n_0$ and $\epsilon$.
For the assumed values of $n_0 = 0.1$\cm3 and $\epsilon = 0.1$, the agreement
is good for the medium-size superbubbles with $r \sim 500$~pc.
The  kinetic energy predicted by the dynamical model is systematically higher,
however, as the shell size increases.  One explanation for this divergence 
is that the average density seen by the expanding shells becomes smaller
as the shells get larger.  
An average density of 0.01\cm3 would bring the values for
the largest shells into agreement.





For an average ambient density of 0.1\cm3, the dynamical model implies 
energy injection rates of $L_W = (10^{-1}(L/n)$\ergsec as shown on the
y-axis of Figure~\ref{fig:stars}.  
The mechanical power available from supernovae and stellar winds, 
solid line, was estimated from the ionizing luminosity of the host galaxy. 
For illustrative purposes, the same conversion from ionizing luminosity 
to mechanical power was applied to all the galaxies (see \S4.1).
Since all the complexes lie below the line, the
current massive star populations in these galaxies could power even
the largest expanding complexes. Unless the star formation rate has increased
dramatically during the lifespan of  a bubble, the mechanical power liberated
by the massive stars is more than adequate to create the observed bubbles.

\subsection{Fate of the Supershells}

Once the massive stars die off, a supershell may in principle continue
to coast outward conserving its momentum.  Eventually, however,
the bubble's environment dictates its fate.  I will consider in turn
the role of radiative losses, the pressure of the ambient medium, and the
gas scale height of the ambient medium.
Since dwarf galaxies typically exhibit solid-body rotation,   
shear forces are expected to be negligible in their disks.

\subsubsection{Cooling Time}

As radiative losses from the bubble's interior become
significant, the shell's growth rate will slow down to that
of the momentum-conserving solution.  In the self-similar solution for the 
shell's motion, the thermal energy in the bubble is 5/11 of the kinetic
energy supplied by the stars, so  in terms of a cooling coefficient
$\Lambda (T)$ the  timescale for the bubble to radiate all its thermal 
energy is
$ t_c \approx (5/11 L_w t ) / (4/3 \pi R_c^3 \Lambda n_H n_e). $
The temperature dependence of the cooling rate  is fit reasonably well 
by  a power law, $\Lambda = \Lambda_0 T^{-0.5}$, between temperatures of 
$10^5$~K and $10^{7.5}$~K.  At cosmic abundance $\Lambda_0$ is
$\approx 1.6 \times 10^{-19}$\ergsec cm$^3$ (Koo \& McKee 1992), which
is probably a good upper limit for $\Lambda_0$ in the hot ISM of the dwarf
galaxies.

The cooling rate is sensitive to the amount of mass residing in the cavity. 
The conventional analytic solution for the temperature and density 
structure inside a bubble is based on classical conductivity theory.
Taking the predictions for the bubble's maximum density and temperature 
as representative of the bubble interior (e.g. Shull 1993),  
the estimated cooling time is 
$t_c \approx (2.46 \times 10^8 ~{\rm yr} ) v_{100}^{1/7} R_{100}^{5/7}
n_0^{-2/7}$. 
Radiative losses from the bubble interior are a concern when the cooling
time is less than or of the same order as the age of the bubble, i.e. when
$ v ~\sles\ (0.51 ~{\rm km~s}^{-1}) R_{100}^{1/4} n_0^{1/4} \kappa_0^{5/8}.$
For reasonable values of the ambient density, 
almost all the shells in \fig~\ref{fig:rv}  lie above this relation.
Based on conventional arguments then, 
radiative losses from the bubble interiors would not
have significantly affected the dynamical evolution of the shells.
However, 
``mass loading'' by evaporation off entrained clouds may drive
the actual mass of hot gas as much as an order of magnitude higher
(Martin \& Kennicutt 1995; Heckman \et 1995; Strickland \et 1996);
and the cooling times may need downward revisions when better \x
data becomes available.

\subsubsection{Pressure Confinement}

The external pressure of the ISM may also suppress a shell's growth.
A bubble is said to be pressure confined  when its internal pressure
falls to that of the ambient ISM (e.g. Koo \& McKee 1992). This condition
is met when the shock speed slows to roughly 13 -- 20\kms.
In \fig~\ref{fig:weaver}, for instance, 
a horizontal boundary at $v \approx 13 - 20$\kms would denote the region where
pressure confinement becomes relevant to a shell's evolution.
All the shells described in \S~\ref{sec:results} are expanding supersonically,
however;  so it is not clear that even the larger shells will stall.

\subsubsection{Breakthrough and Blowout}

As a shell outgrows its galaxy, the flattening of the ambient gaseous
disk is expected to collimate the outflow along the minor axis
(e.g. DeYoung \& Heckman 1995). A shell will breakthrough the disk
if it is expanding supersonically when it reaches one gas scale height.
Puche \et (1992) measured an HI scale-height of 625~pc in Ho~II ,
so the gas scale height in dwarf irregulars
may typically be larger than that of late-type spirals.
For a  midplane
gas density $n_0$, effective scale height $H = 500$~pc, and sound speed
$c_0 = 20$\kms, the critical energy injection rate for breakthrough
is 
$L_{crit} / n_0 \ge 8.0 \times 10^{38} H_{500}^2 c_{20}^3$
\ergsec cm$^3$.  In \fig~\ref{fig:weaver}, 
the thick dashed line illustrates this critical power for breaking through
the disk.  Many of the supershells lie above this relation, so
the shells are expected to breakthrough the HI disk.

Maps of the HI 21-cm emission support this argument.
The HI distribution in three dimensions is often
not uniquely determined for dwarf galaxies, but
the position angles of the polar shells in \fig~\ref{fig:images} are
preferentially oriented closer to the minor axis of the HI emission than
the major axis.  The lack of perfect alignment indicates
the collimation occurs on smaller spatial scales.
The dimensions of the HI emission, typically at 
the $10^{19}$ or $10^{20}$cm$^{-2}$
column density contour, are given in Table~\ref{tab:sam}.
In \n1569, \n1800, \n2366, IZw18, M82, \n3077, \n4861, and \n5253 
the polar bubbles have grown to heights beyond much of the HI disk.  
The same may be true in \n3738, \n4214, and \n4449, although the low
inclination precludes direct observation.  Although the prevalence
of very low-density HI halos is unknown, Sex~A is the only galaxy
in the sample with a reported detection.
Hence, it would appear that
these polar shells will meet little more resistance from the ambient	
HI disk/halo. 

An important question is whether the shells rupture opening a conduit
for the escape of the hot gas in their interior.
Numerical simulations suggest the shells do not accelerate and fragment 
from Rayleigh-Taylor instabilities until they reach several scale heights
(Mac Low, McCray, \& Norman 1989).  The minimum
energy injection rates for blowout are therefore ``several'' times
higher than for breakthrough alone.  Blowout still seems likely, however, 
for many of the bubbles in \fig~\ref{fig:weaver} lying above 
$L/n = 10^{40}$\ergsec\ cm$^3$.
The fate of this material then depends on the escape velocity several
hundred parsecs above the HI disk.


\section{Mass Loss}
\label{sec:mdot}

Figure~\ref{fig:images} shows numerous examples of warm ionized
gas being lifted out of the galactic disk.  To determine whether this
material will escape from the  galactic gravitational potential,
a model of the distribution of the visible and dark matter is 
used to calculate the escape velocity.  Low-resolution HI maps and rotation 
curves have been published for most of the galaxies in this study.  
Beam-smearing artificially flattens the rise of these rotation curves, so it 
would be misleading to describe {\em fits} to the rotation curves.  The  data 
do, however, constrain the maximum circular velocity and provide a measure of
the Keplerian mass  within that radius $r(v_{\rm circ})$ (Table~\ref{tab:sam}).
In this section I examine this constraint on a galaxy by galaxy basis
and discuss the normalization of the dark (hereafter {\em halo}) component.  
Assuming the halo density profiles resemble the  profiles derived empirically
from low surface brightness galaxies (Burkert 1992),
the escape velocity from dark halos extending to roughly the virial radius 
is easily computed.


\subsection{Galactic Mass Loss? Case Study NGC~1569}



HI rotation curves for \n1569 from Reakes (1980) and  Israel \& van 
Driel (1990) are reproduced in \fig~\ref{fig:rotate}.  
The circular velocity at $r = 1$~kpc is $\sles\ 31$\kms.
The {\rm projected} expansion speeds of the kiloparsec-scale shells are
1.4 to 4.0 times higher, so \n1569 has the largest ratio of shell speed 
to maximum circular velocity in the sample.  It is perhaps the most likely 
of any of the nearby starbursts to lose a substantial fraction of its
warm ionized gas mass to the intergalactic medium, so 
the limits on its escape velocity are particularly interesting.

The dashed and dotted lines in \fig~\ref{fig:rotate} show
the rotation curves for dark halos with core radii of $r_0 = 0.53$~kpc
and 1.4~kpc, respectively.  This one-parameter family of models
is a good description of the dark matter distribution in the seven low 
surface brightness galaxies where it has been directly measured (Burkert 1995).
The circular velocities of these Burkert halos are significantly lower than 
observed if one accepts either the turnover in the Reakes rotation curve at 
$r \sim 1$~kpc or the higher velocities in the Israel \& van Driel (1990) data.

The distribution of the {\em extra} mass in the inner galaxy significantly 
affects the extrapolated mass distribution at large radii.
The resolution of the HI rotation curve is not sufficient to distinguish
between a denser halo model and a disk-like matter distribution,
but a maximum disk model can be compared to the visible mass in the disk.
For example, the circular velocity in a disk with an exponential surface 
density profile and a scalelength similar to the B-band surface brightness 
profile, $\alpha^{-1} \approx 530$~pc, would reach a velocity of 
$\sim 34$\kms at 1.1~kpc for a central surface density of 240\msun/pc$^2$.  
(The central density can be reduced if the rotation curve continues to rise and 
the disk scalelength is increased.) The B-band surface brightness within one 
effective radius is 39\lsun~pc$^{-2}$ (de Vaucouleurs \et 1991).
For a purely exponential surface
brightness profile, the central surface brightness would be about 5 times
higher or $\mu_0 \approx 200$\lsun~pc$^{-2}$.
Although the stellar mass to blue light ratio of the $\sim 10$~Myr old burst
is very small ($\sim 0.1$), even a small ($\sgreat\ 10\%$) contribution to the
light from an old (i.e. several Gyrs) population would raise the integrated
$M/L_B$ ratio to $\sgreat\ 1.1$ (Bruzual \& Charlot 1993).  In addition,
for  $(M/L_B)_* \sim 1 \msun / \lsun$, the atomic gas mass,
$1.4 M_{HI}$, or $\sim 1.2 \times 10^8\msun$, exceeds the stellar mass,
$L_B = 1.62 \times 10^8$\lsun. Some of this material lies beyond a radius of 
1~kpc, but clearly the  Keplerian mass  interior to 1~kpc 
($\sim 2.2\times 10^8$\msun) can be accounted for without invoking
a dark matter component.

\n1569 would be  very unusual, however, if it did not have a dark matter halo;
and the dominant contribution from luminous, disk mass at radii less than 1~kpc
in no way rules out the presence of a large, diffuse halo. 
Adding an $r_0 = 1.4$~kpc Burkert halo to a disk model with scalelength
500~pc, the central density of the latter is only reduced to 
155\msun\ pc$^{-2}$.  Even a halo like that around DDO~154, core radius 
$\sim 2.9$~kpc, would contribute only $7.8 \times 10^7$\msun\
within 1~kpc, although it would cause the rotation curve to turnover at
a much larger radius than observed.  A substantial range of composite
disk plus halo mass distributions are then consistent with the data.
The escape velocity from one of them is 
$v_{esc} = \sqrt{2} \sqrt{|\phi(R,z)|}$,
where the galactocentric radius $R$ is equal to zero above
the burst region and $\phi = \phi(disk) + \phi(halo)$.
The halo potential is the integral of the gravitational force on 
a test mass from radius $z$ to the halo truncation radius.
The potential at a height $z$ above the center of an infinitely thin disk 
can be computed using Toomre's method to solve the
Laplace equation (Toomre 1963; Binney \& Tremaine 1987 \S2.6).



Figure~\ref{fig:potent} compares the escape 
velocity above the galactic center for three models of the mass distribution.
The exponential disk model ($\alpha^{-1} = 500~{\rm\ pc}, \sigma_0
= 240$\msun\ pc$^{-2}$) with no halo is a lower limit on the escape velocity.
A more realistic model, dotted line, includes a spherically symmetric
Burkert halo  with core radius $r_0 = 1.4$~kpc and truncation radius 160~kpc.
Lowering the central disk density to 155\msun pc$^{-2}$ to maintain
a circular velocity of $\sim 30$\kms at $r = 1$~kpc, the
escape velocity from the composite potential is  raised to almost $100$\kms.
The third model has equal mass contributions from the halo and disk 
components inside 1~kpc.  Since the halo central density must be increased by 
a factor of 1.5 over Burkert's $\rho_0(r_0)$ relation, this model can
be thought of an upper limit on $v_{esc}$. 

The expansion speeds of the polar shells in \n1569 are shown as
filled circles in \fig~\ref{fig:potent}.  
The  shell velocities range from 40\kms to 120\kms. Several are 
expanding faster than the escape velocities from the disk model at 
$z = 1$~kpc. When a halo component is included, however, the expansion of
all but one of the shells is sub-critical.  Since the {\em projected}
expansion speeds are $\sles\ v_{esc}$, these shells are marginally bound.
In other words, some shell mass could permanently  escape -- i.e.
corrections for projection effects could easily increase $v$ by 50\%, but
much of the warm ionized gas probably remains bound to \n1569.





\subsection{Escape Velocity Estimates}


As illustrated in \fig~\ref{fig:vall}, the resolution of the HI rotation
curves for the other galaxies varies from good, i.e. Sextans~A,  to 
essentially a single element, i.e. \n4861. 
Inclination corrections, see Table~\ref{tab:sam}, have been applied, but
the low resolution and non-circular gas motions may introduce
significant errors in the Keplerian masses derived for a few objects.
The descriptions below provide a qualitative impression of the possible 
magnitude of such systematic errors.  Array data have not been
published for \n2537 and \n3738, and the three galaxies without superbubble
detections are not shown.  

The Keplerian mass  provides some guidance for the potential models.
The difference between the dynamical mass and the mass of atomic gas in 
Table~\ref{tab:mdot}, for example, can only be explained with a stellar 
component if its mass to light ratio is as high as that in Column~8.
The ratios for \n1569, \n3077, \n4214, and \n5253 are fairly typical for 
stellar populations, so  a baryonic disk could dominate the potential in 
the inner few kiloparsecs. In the other dwarfs, however,
some form of dark matter apparently dominates the mass in even
the regions interior to $R(v_{circ})$. 
The spatial distribution of the matter is unknown, 
and a disk of cold molecular gas could contribute much of the mass. 
To place a conservative upper limit on the escape velocity, however,
the extra mass is best modeled as an increase in the central density of 
the dark halo.

\subsubsection{Sextans A}
 
Sextans~A is one of only three sample members with $M_B > -14$~mag
and the only member of the Local Group. 
The rotation curve along the kinematic major axis (Skillman \et 1988)
is shown by the squares in \fig~\ref{fig:vall}.
The small ionized shell has an expansion velocity, $\sim 60$\kms,
significantly higher than the rotational speed of the neutral gas at
the galactocentric radius of the first-ranked HII region.
The elliptical orbits of the HI in the inner galaxy have been interpreted
as evidence for  a gaseous bar (Skillman \et 1988), and the first-ranked
and second-ranked HII regions are located near local maxima in the
HI column at opposite ends of this bar.
Dotted lines illustrate the rotation curve and escape velocity of
an $r_0 = 532$~pc Burkert halo.  
An increase of a factor of 5.3 in the halo central density produces
a better fitting rotation curve, solid line.


\subsubsection{IZw18}
One might expect mass loss of catastrophic proportion, since IZw18
is one of the least luminous galaxies yet contains a kiloparsec-scale
polar bubble.  (Note that only the southern lobe of the bipolar shell 
discussed by Martin (1996) was detected kinematically and therefore cataloged 
Table~\ref{tab:bubbles}.)
If the HI velocity field actually reflects a rotating
disk in equilibrium, however, the dynamical mass within $r = 850$~pc
is $\sim 5 \times 10^8$\msun (Viallefond \et 1987; van Zee \et 1998)
 -- a bit more than \n1569!  
In \fig~\ref{fig:vall}, the projected expansion speed of the polar shell,
30\kms, is less than the escape speed from an $r_0 = 1.4$~kpc Burkert halo
(dotted line). 
The deprojected shell velocity, closer to $\sim 90$\kms (Martin 1996),
may exceed this escape velocity. 
The central density
must be increased by $9.4$ over Burkert's $\rho_0(r_0)$ relation to
reach a circular velocity of 50\kms (solid line).
Since the dynamical mass estimate greatly exceeds the mass of this halo
model, the argument for escape of the warm ionized shells is marginal.

\subsubsection{NGC 1800}


The shallow rise of the rotation curve interior to $r = 6$~kpc is not
entirely a resolution effect.  In the inner part of \n1800,
the HI velocity gradient rotates by 90\deg\ and lies along the galaxy's
minor axis (Hunter, van~Woerden, \& Gallagher 1994).  Hunter \et (1994)
suggested the  gas is streaming along a stellar bar, and  the 
ellipticity difference between the isophotes and isochromes 
confirm that the potential is non-axisymmetric (Quillen \et 1996). 
  The dynamical mass estimate is therefore
uncertain since the  gas is probably not on circular orbits in the inner 
few kiloparsecs, and the HI cloud about 2\farcm4 west of the galaxy may
not be in a circular orbit either.  At a radius of 1.3~kpc the escape 
velocity from an $r_0 = 1.4$~kpc halo,  about 70\kms,  is somewhat higher 
than the expansion speed of the northern polar bubble, 50\kms.
Increasing the central halo density
by a factor of 1.9 to fit the circular velocity at $r = 4.7$~kpc
raises the escape velocity to nearly 100\kms.

\subsubsection{NGC 2366 and NGC 2363} 

\n2363 is a giant HII complex $\sim 90\asec$ southwest of the
dynamical center of the dwarf galaxy \n2366 (e.g. Braun 1995).
Since the distance of \n2363 from the center of the potential
well, $R \sim 1.6$~kpc, is greater than the radius of the bubbles,
the shell velocities are plotted in \fig~\ref{fig:vall} against the
radius of \n2366 rather than the shell radius. 
The expansion speed of shells~A and B is less than   the
escape velocity, $\sim 145$\kms, from an $r_0 = 3.76$~kpc Burkert halo.
This halo model provides a reasonably good description of the rotation
curve, so the escape velocity is unlikely to be much higher.

\subsubsection{NGC 3077}

The HI distribution in \n3077 is asymmetric.  An HI stream extends
northward from a large concentration in the southeast toward the outer
spiral arm structure of M81 (Appleton \et 1981; van der Hulst 1979).
Despite this interaction, the velocity gradient along the optical major
axis is surprisingly regular.  The rotation curve in \fig~\ref{fig:vall} 
was estimated from the position -- velocity diagram along the optical 
major axis (van der Hulst 1979) and corrected for an inclination of 38\deg.
The maximum circular velocity at $r = 6$~kpc is similar to 
the velocities of shells \n3077-- A, J, B, and G, so the shells
are traveling slower than the escape velocity. The bold, dashed line 
shows the escape velocity above the center of a disk with scalelength 3.76~kpc
 and central surface density 133 \msun~pc$^{-2}$.  The escape velocity
from an  $r_0 =  3.76$~kpc Burkert halo is similar.  These models
have a mass of $5.9 \times 10^9$\msun\ within $R(v_{rot}) = 6.3$~kpc.
Higher resolution data might allow a smaller scalelength and lower escape
velocity.


\subsubsection{NGC 4861}

The giant HII region shown in \fig~\ref{fig:images} is about 1\amin
(2.18~kpc) southwest of the dynamical center of \n4861 (e.g. Wilcots
\et 1996b).  The HI iso-velocity contours
increase by 80\kms from south to north
over about 15\farcm\  The circular velocity is then $\sgreat\ 43$\kms at a 
radius $r \approx 16.5$~kpc.  The expansion speeds of 
shells A, B, and~C are plotted against the distance of the giant
HII region from the dynamical center of the galaxy in \fig~\ref{fig:vall}.
The escape velocity from a Burkert halo with core radius $r_0 = 2.31$~kpc
is shown for comparison. The expansion speeds of shells A and B 
are of the same order as the circular velocity.
The escape velocity at their location is probably a factor of several
higher.

\subsubsection{NGC 5253}

The relatively high resolution (FWHM~$\approx 10\asec$) HI map obtained by
Kobulnicky (1997) shows a large disturbance in the radial velocity field
southeast of \n5253.  The velocity gradient along this half of the galaxy's
minor axis is greater than that along the major axis.  Similar gradients
were measured in the \Ha velocity field (Martin \& Kennicutt 1995).
Hence, rotation does not dominate 
the velocity field of the interstellar gas in \n5253.
The mass of visible gas and stars does place a lower limit on the 
dynamical mass of \n5253.  For a composition H/He = 10/1 by number,
the mass of neutral gas is $\approx 1.96 \times 10^8$\msun. The
stellar mass is $\sim 2.3 \times 10^8$\msun\ for an assumed 
$(M/L_B)_* = 1.0 \msun /\lsun$.  The Keplerian velocity at 
the extent of the detected HI, $r = 1.24$~kpc,
would then be at least 38\kms.  If 
the mass is centrally concentrated, then the circular velocity
at $r = 500$~pc, the edge of the \Ha nebula, could be as high
as 60\kms.  The estimated escape velocity is at least a factor 
of 1.4 times higher.  The swept-up shell 
expanding at $\sim 35$\kms is then marginally bound to \n5253.

\subsubsection{NGC 4214}

The HI rotation axis in \n4214 is nearly aligned with the optical
major axis; and McIntyre (1996) suggests 
the strong central iso-velocity twist could be the signature of a bar.
The extended disk of HI in \n4214 reaches a circular rotation speed
of $\sim 70$\kms at a radius $R= 2.91$~kpc (McIntyre 1996).
Like \n3077,  much of the mass in the inner region of \n4214 could, however,
be provided by a disk  of stars and gas. The bold, dashed line in the figures 
shows that the escape velocity above the center of a disk with central surface
density 235 \msun/pc$^2$ and scalelength $r_0 = 2.91$~kpc.
The escape velocity from an $r_0 = 2.91$~kpc halo model is lower, but that
from a denser halo that dominates the interior mass would be higher.
 Both estimates of the halo escape velocity one 
kiloparsec from the starburst are somewhat   higher than the shell speeds 
which range from 28 -- 100\kms.


\subsubsection{NGC 4449}

The HI surrounding \n4449 extends about 14 times farther than the
optical galaxy. The gravitational interaction with DDO~125 is probably
responsible for some distortion in the velocity field
(Bajaja, Huchtmeier, \& Klein 1994), but the disk seems to be in
regular rotation about the center of \n4449 (Wilcots \et 1996a).
The velocity increases from north to south across the 
extended disk -- a gradient opposite
to that observed across the inner few arcminutes (e.g. see
echellograms \n4449--1, 2, 7, 
and 10).  The rotation curve in \fig~\ref{fig:vall} is based on
the position -- velocity diagram that Bajaja \et extracted along the
HI ridgeline.  The  velocity rises to 65\kms
15.7~kpc from the galactic center.
For an inclination of 51\deg,  the maximum circular velocity is then
84\kms which is near the middle of the range of shell expansion speeds
(30 -- 150\kms).  For comparison, the circular velocity of a halo
with core radius $r_0 = 6.3$~kpc is shown in \fig~\ref{fig:vall}.
If a spherically symmetric halo of this type is a good
description of the gravitational potential in \n4449, then the gas
swept into the expanding shells is bound to \n4449.

\subsubsection{M82}

The last panel
\fig~\ref{fig:vall}  shows a rotation curve  derived from measurements
of stellar absorption lines along the major axis of M82 (Burbidge,
Burbidge, \& Rubin 1964).
The light and bold dotted lines show the circular velocity and escape velocity,
respectively, from an $r_0 = 3.76$~kpc Burkert halo.  Increasing the
central density by a factor of  $22.1$ over Burkert's 
$\rho_0(r_0)$ relation to raise the interior halo mass to the estimated
dynamical mass raises  the upper limit on the escape velocity from 130 \kms
to 650\kms.  Although this outflow is clearly in a differenct class
energetically, it is not clear that even this warm ionized gas will
escape the deeper potential.

\subsection{Conclusions}

\subsubsection{Galatic Mass Loss and Blow Away}

The circular velocities of the program galaxies are compiled in 
Table~\ref{tab:sam} for reference. They span a factor of 4 in magnitude -- a 
range similar to the velocity distribution of kiloparsec-scale shells.
The observed mass of stars and gas alone place a substantial lower limit on
the escape velocity in several galaxies. Halos with core radii that reproduce
the turnovers in the rotation curves predict even larger escape velocities since
they extend to $\sim 160$~kpc. The best estimates of the  escape speeds from 
the current data are typically a factor of 1.5 to 3 times
higher than the projected expansion velocity.  
However, while the potential does appear to be relatively  deep in
some of the galaxies with the most energetic outflows (e.g. M~82 and \n4449),
large and/or fast shells are also found in some less luminous galaxies. 
In the case of Sex~A, and quite possibly I~Zw~18, the shell velocities are 
comparable to $v_{esc}$.  
Substantial loss of shell material seems most likely in \n1569 where
mutiple shells have propagated well above the galactic plane, but the
galaxy is apparently not very massive.

In contrast to the warm ioninzed gas in the shells, the hot, \x emitting gas at 
high latitudes is not bound to the galaxy.  An escape velocity of 100\kms, for
example, is equivalent to an escape temperature of $4 \times 10^5$~K
(e.g. Wang 1995; Martin 1996).  The soft \x halos detected around
\n1569, \n4449, and M82 have \x temperatures of $ \sim 7 - 9 \times 10^6$~K
(Della Ceca \et 1996; Della Ceca \et 1997; Strickland \et 1997).  
The position angles of the
\x lobes  protruding from these galaxies correlate strongly with
those of the extended \Ha filaments, so the lobes of hot gas are probably
at least partially confined by cooler surrounding gas.  
This hot gas will escape from the galaxy in a wind
if the confining shell  breaks up.  Estimates of the mass in the
hot ISM are compiled in Table~\ref{tab:mdot} for a volume filling factor
$\sim 1$.  The mass in hot gas is 
only a few percent, of the interstellar HI mass in the two dwarfs galaxies
where it is measured.  

The current star formation rates in these galaxies are at least
an order of magnitude too low to instantly (i.e. $t < 10^7$~yr) expel the 
ISM.  For each galaxy with a dynamical mass estimate, the total kinetic energy in 
large-scale shells was compared to a lower limit on
the gravitational binding energy of the interstellar gas,
$E_B \sgreat\  (1.4 M_{HI}) v_{circ}^2$.  Their ratio, column~11 of 
Table~\ref{tab:mdot}, is less than one for all the galaxies.  The
\n1569 outflow has the highest ratio, $K.E. / E_B \sles\ 0.3$.
As emphasized by DeYoung \& Heckman (1994),  the transfer
of this energy, which is largely in polar outflows,  to the ambient disk
may be quite inefficient.  Hence, if the star formation rate is rapidly
declining in these galaxies, then the bulk of their interstellar gas
will remain in the disk.

\subsubsection{Disk Mass Loss}

These data demonstrate that superbubbles lift significant masses of
gas out of the star-forming disk. Regardless of whether any shell material actually 
escapes from the galaxy, the efficiency of this {\em re-heating}
is an important parameter in models of galaxy formation and evolution.
Table~5 summarizes the upper limits on the {\em disk} mass loss rates.
These rates represent the transfer of warm ionized gas 
and hot ionized gas from the star-forming disk 
to the halo and were calculated by
dividing the mass in ionized shells
and hot gas by the dynamical age of the oldest polar bubble.
Although measurements of the filling factor $\epsilon$ may lower the
mass loss rates by factors of a few, the disk mass loss rates
will likely remain significant as they are typically slightly greater than
the current star formation rates.

If star forming regions similar in scale to the current
giant HII regions continue to form, the mass blown into the halo of 
several of the galaxies will be significant.
The timescale for removing the ISM  is  of order
 the total gas mass, $\sim 1.4 M_{HI}$, divided by the
estimated mass loss rates in Table~\ref{tab:mdot}.
These values range from 50~Myr and 140~Myr for M82 and \n1569,
respectively, to 11~Gyr for \n4214.  
In Table~\ref{tab:mdot}, I compare these timescales to
the rotational period of the disk.  The rotation period was computed
at the radius where the circular speed was tabulated and is representative
of the inner disk which shows solid body since rotation.
Associating this period with a timescale for star formation, $\tau_*$, the
bulk of the HI disk could be removed in less than $6 \tau_*$
for the following galaxies: \n1569, \n1800,  M82, \n3077,
\n4449, and \n4861.




\section{Summary}
\label{sec:sum}

An extensive set of \Ha echellograms and images were used to reconstruct the
large-scale kinematics of the ionized gas in 14 dwarf galaxies and
M82.  Details of the  results for individual galaxies are
included in their respective subsections of the  paper, and
\fig~\ref{fig:weaver} provides a concise summary of the shell
expansion speeds and sizes.  The main results
regarding the formation of winds in dwarf galaxies
are summarized here.

\begin{itemize}

\item
The formation of supershells must be a common byproduct of massive
star formation in dwarf galaxies. 
Expanding, supergiant ($R > 300$~pc) shells were found in all but
two of the galaxies.  This sample was drawn from the population of
nearby dwarf galaxies with prominent  arcs and/or extended filaments
in their \Ha emission, and roughly one out of four catalogued dwarf
galaxies fits this description (Hunter, Hawley, \& Gallagher  1993).
Indeed, the hierarchical growth of these structures
probably began in star forming regions akin to 30~Doradus in the Large
Magellanic Cloud (Chu \& Kennicut 1994).
The most powerful  outflows, i.e. \n1569 and M82, were found
to be composed of multiple {\em cells} whose walls are
probably the interface between polar shells. Star formation in
the  lowest luminosity galaxies, e.g. IZw18, also generates
kiloparsec-scale shells.

\item
Although many of the expanding complexes survive for 10~Myr,
none older than $\sim 20$~Myr were identified.   The lack
of shells older than this likely reflects their disruption timescale
and provides an indirect measure of the scale height of the ISM.
Alternatively, the ionization rate of the shells might drop abruptly
on this timescale due to changes in the birthrate of massive stars
and/or the illumination geometry.  Although
bright, extraplanar HI shells have not been detected in any of the
galaxies in this sample, some HI holes in the LMC (Meaburn 1980) and 
\n4449 (Hunter \& Gallagher 1997) are probably relics of expanding supershells.
The power requirements of the ionized supershells typically exceed
the critical power for supersonic disk breakthrough, so 
a disruption scenario must be favored for them. 
The sequence of echellograms stepped across the southern lobe of 
\n1569 constrains their deceleration and
shows multiple velocity components at least up to 640~pc above the galactic
plane.
Future observational work must aim to detect the remains of the
hot gas and ruptured shell following blowout.  One might speculate
that the quiescent filaments in \n5253 could be fragments of a ruptured
shell or clumps of infalling material ejected in a previous wind
epoch.  A better census of the local dwarf population would also be
helpful for constraining the duty cycle of the winds.


\item
Presuming the shells do rupture, the escape of  hot, X-ray
emitting gas from their interiors seems certain. A diffuse,
thermal component of the \x emission has been resolved in three
of the galaxies in the sample, but it is  only a significant fraction 
($> 10\%$)  of the interstellar HI mass in M82.  In contrast,
much of the interstellar gas swept into the warm ionized shells probably
remains bound to the galaxy. The structure of the dark matter halo has been 
measured in several low surface brightness dwarf galaxies with large HI disks
(e.g. Carignan \& Beaulieu 1989; Meurer \et 1994) and appears to have
a universal structure (Burkert 1995; Navarro, Eke, \& Frenk 1996).  
Hence, a conservative approach to mass loss is to assume that the bursting
dwarfs are embedded in similar halos.  These dark halo models often do not 
provide enough mass to explain the HI rotation speed in the inner galaxy, 
however. Stars and atomic gas can account for essentially all the dynamical 
mass inside $R(v_{circ})$ in some galaxies like \n1569, but the fraction 
varies enormously among the sample members.  Although little CO emission is 
detected from the dwarf galaxies (e.g. Young \et 1996), the large uncertainty 
in the $H_2$ to CO  conversion factor does allow substantial mass contributions
from molecular gas (e.g. Maloney \& Black 1988). A dominant disk in the 
inner few kiloparsecs has two immediate implications.
First, the disk mass contributes significantly to
the gravitational acceleration of the kiloparsec-scale shells/disk outflows.
Second, the observed turnover in the rotation curve may not
be revealing much about the core radius of the halo -- a critical parameter
for estimating the escape velocity.
The maximum circular velocities in the galaxies do generally appear to be
comparable to the expansion velocities of the supershells, so
the escape velocities are greater than  the projected shell speed.
The expansion speeds along the minor axis of \n1569 do
reach values several times the maximum rotation speed.



\item
The warm shells alone lift gas out of the disk at rates comparable to or 
greater than the current  galactic star formation rates.
The shells transport $10^5$ to $10^6$\msun\ of gas over kiloparsec-scale
shells in 10~Myr and leave the sound speed high in a large volume of the ISM.
This hydrodynamic mixing will be faster than a diffusion process, so
the bubbles will clearly alter the chemical evolution of these galaxies.  
The timescales for blowout are shorter than the evolutionary timescales of most
models for Type~Ia supernova progenitors, so the mass loss begins 
before much of the iron from the burst has been mixed into the ISM.
The composition of the ejected material will depend on the duration of the 
wind and the composition of the ambient ISM, so their impact on the galactic 
chemical evolution is interwoven with the galactic star formation history. 

\item
Although the current kinetic energy in the large
expanding structures is only comparable to the binding energy of the ISM  
in \n1569, bubble blowout may still extinguish
star formation in  particular regions of the other galaxies.
If the hot spots percolate across the dwarf irregular galaxies for many
rotational periods, then a substantial fraction of the interstellar gas
may still be cycled through a halo and/or lost from the galaxy.
If the present mass loss rates could be sustained
for 6 orbital timescales, for example,
then most of the interstellar HI could be removed from the disks of
six of the 15 galaxies.  
This global gas-dynamical feedback will be discussed in the context of
the galactic star formation history in a forthcoming paper.


\end{itemize}

\acknowledgements{I would like to thank
Chip Kobulnicky, Vince McIntyre, and Eric Wilcots for 
providing results from HI observations prior to publication.
I am grateful to Rob Kennicutt and Tim Heckman for 
their comments and suggestions about the manuscript, and to
Rob Kennicutt for assisting with some of the initial data
acquisition.
Financial support was provided by an NSF Graduate Fellowship,
Hubble Fellowship grant HF-01083.01-96A, and NSF grant AST-9421145.

\newpage
%
%

\newpage
\begin{table}[h]

\caption{The Sample of Galaxies}

\begin{tabular}{lllllllll}
\tableline
\tableline
Galaxy 	&  d\tablenotemark{a}
       	& $M_{B}$\tablenotemark{b}
       	& D\tablenotemark{c}
	& HI Extent\tablenotemark{d}
        & $v_{circ}^{proj~}$\tablenotemark{e}
        & $v_{circ}$\tablenotemark{f}
        & $R(v_{circ})$\tablenotemark{g} 
	& HI Ref.\tablenotemark{h} \\
       &(Mpc)& (mag)   & (\arcmin) 	&(kpc $\times$ kpc) & 	(km/s) 	  & (km/s)   & 	(kpc)	     &   \\
\tableline		                    		               			
VII~Zw~403 & 3.6&	 -13.68&3.6	&\nodata            &    	\nodata	  & \nodata  & 	\nodata	    & 	 1\\
Sextans~A   & 1.31 &	-13.84 &1.31	&3.4 $\times$ 3.4   & 	20	  & 34   &	1.2      & 	 	2,3\\
I~Zw~18  & 10.0 &	-13.84 &0.5	&2.9 $\times$ 1.5   & 	50	  & \nodata &  	0.853      & 	 	4,5\\
II~Zw~40 & 10.1 &	-14.54 &1.3	&15.2 $\times$ 2.4  & 	53	  & \nodata   &	5.2  	     & 	 	6,7\\
NGC~3738&  4.0    &-16.0&4.0	&\nodata            &    	\nodata	  & \nodata &  	\nodata	    & 	 8\\
NGC~1800& 8.1  &	-16.67 &2.2	&13.9 $\times$ 8.08 & 	40	  & 46        &	5.7  	     & 	 	9,10\\
NGC~2363& 3.6  &	-16.75 &8.0	&7.67 $\times$ 16.7 & 	53	  & 63        &	5.5       & 	 	6,11\\
NGC~1569& 2.2  &	-17.26 &3.1	&4.2 $\times$ 3.2 & 	28	  & 31	      &	1.0       & 	 	12,13\\
NGC~2537& 7.5  &	-17.36 &1.7	&\nodata            &    	\nodata	  & \nodata  & 	\nodata  &	 6\\
NGC~3077& 3.6  &	-17.37 &3.6	&12.7 $\times$ 8.0  & 	40	  & 65   &	6.3      & 	 	14,15\\
NGC~4861&  7.5 &	-17.6  &7.5	&11.9 $\times$ 15.7 & 	40	  & 43        &	16.5      & 	 	16\\
NGC~5253&  4.1 &	-17.62 &4.1	&4.3 $\times$ 4.3   & 	$<15$	  &  38 - 60  &	1.24      & 	 	17,18,19\\
NGC~4214&  3.6  &-17.65 &3.6	&14.6 $\times$ 12.9 & 	35	  & 70   &	2.91       & 	 	20,21\\
NGC~4449&  3.6 &	-17.86 &3.6	&43 $\times$ 70     & 	65	  & 84        &	15.7      & 	 	22 \\
M~82    & 3.6  &	-18.95 &3.6	&4.19 $\times$ 1.57 & 	135	  & 136       &	2.0       & 	 	14,23,24\\
\tableline	
\end{tabular}

\scriptsize
\tablenotetext{a}{(col. 2) Adopted distance.}
\tablenotetext{b}{(col. 3) Absolute blue magnitude from the RC3 blue magnitude (deVaucouleurs 1991). }
\tablenotetext{c}{(col. 4) Angular diameter at the 25 B-magnitudes per square arcsecond 
isophote from Tully (1988). }
\tablenotetext{d}{(col 5) Spatial dimensions of the neutral hydrogen. }
\tablenotetext{e}{(col 6) Projected rotation speed. }
\tablenotetext{f}{(col 7) Rotation speed corrected for the disk inclination when
known. }
\tablenotetext{g}{(col 8) Radius at which the rotation speed in col.~6 was measured. }
\tablenotetext{h}{References: 
(1)  Tully,  Boesgaard, Dyck, \& Schempp 1981.
(2) Skillman, Terlevich, Teuben, \& van Woerden 1988.
(3) Huchtmeier, Seiradakis, \& Materne 1981.
(4)  Viallefonde, Lequeux,  \&  Comte 1987.
(5) Lequeux, \& Viallefond 1980.
(6) Thuan, \& Martin 1981.
(7)  Brinks, \& Klein 1988.
(8) Hunter, Gallagher, \& Rautenkranz 1982.
(9) Hunter, van Woerden, \& Gallagher 1994.
(10) Gallagher, Knapp, \& Hunter 1981.
(11) Braun 1995.
(12) Israel \& van Driel 1990.
(13) Reakes 1980.
(14) Appleton, Davies, \& Stephenson 1981.
(15) van der Hulst 1979.
(16) Wilcots, Lehman, \& Miller 1996.
(17) Kobulnicky, \& Skillman 1995.
(18)  Reif, Mobold, Goss, vanWoerden,  \& Siegman 1982.
(19) Kobulnicky 1997.
(20) Kobulnicky \& Skillman 1996.
(21) McIntyre 1997.
(22) Bajaja, Huchtmeier, \& Klein 1994.
(23) Burbidge, Burbidge, \& Rubin 1964.
(24) Yun, Ho, \& Lo 1993.}

\normalsize

\label{tab:sam}
\end{table}
\newpage

\begin{table}[h]

\scriptsize

\caption{Observations \label{tab:obs}}
\begin{tabular}{lllll}
\tableline
\tableline
Telescope 	& Instrument/Detector	&	Date	&	Filter/Grating	& Conditions	\\
\tableline
Steward 2.3 m& Cass Camera + Loral 800 $\times$ 1200 CCD 
	& 1993 April 21 and 28
		& \line\ 6450, \line\ 6580, 70\angs\ FWHM 
& Clear		\\
``		&''			& 1993 May 16-17&	``		&  ``		\\
``		&''			& 1994 January 3&	``		&	`` 	\\	
``		& Cass Camera + 2k $\times$ 2k CCD & 1993 December 5--6		& ``	& Variable \\
``		& Focal reducer + TI $800 \times 800$	   & 1990 March 22		&''		&  \\
KPNO 4m		& Echelle + Red Long + T2KB	& 1994 April 28 - May 2 & \Ha 6563B, 79 - 63\deg  & Good\\
``		& ``				& 1995 Dec 10 - 11, 1996	&   ``		& Variable \\
``		& ``				& February 7 - 8 	        &  ``		& `` \\
\tableline
\end{tabular}

\end{table}

\newpage
\begin{table}[h]

\vspace*{-1.0cm}

\tiny

\caption{Properties of Supershells \label{tab:bubbles}}
\begin{tabular}{rrrrrrrrrrr}
\tableline
\tableline
Complex~\tablenotemark{a}
	& Slits~\tablenotemark{b}
	& v~\tablenotemark{c}
	& R~\tablenotemark{d}
	& R 
	& Aperture~\tablenotemark{e}
    	& $\Sigma_{{\rm H}\alpha+[NII]}$~\tablenotemark{f}
        & M$\sqrt{\epsilon/0.1}$~\tablenotemark{g}
	& KE $\sqrt{\epsilon/0.1}$~\tablenotemark{h}
	& $\tau$~\tablenotemark{i}
   	& $L_{in}(1)/\bar{n}(1)$~\tablenotemark{j}   \\
	&	&(km/s)&	(\asec)	&(pc)& (asec$^2$) &	  & ($10^6$ \msun) &($10^{51}$ ergs) 		&
(Myr)	&	 (cm$^3$ ergs/s)	\\
\tableline 
SexA-A	  &1    &	62	&13.9   &88    & 7.209e2	&2.38e-16	       &0.027 &	1.0  &0.83	&7.3e38 \\ 
IZw18-S	  &1    &	34	&19.9   &970   & 1.620e2	&1.98e-16	       &0.47 &	5.4  &17	&1.5e40 \\	  
NGC3738-A &2,3 &	35	&8.9    &173   & 1.878e2	&1.58e-15	       &0.060 &	0.73   &2.9	&5.1e38	 \\
	B &2    &	24	&6.8    &132   & 9.027e1	&2.43e-16	       &0.015 &	0.088   &3.2	&9.6e37	 \\
	C &2   &	17	&14.9   &288   & 2.535e2	&9.47e-16	       &0.13 &	0.38   &9.9	&1.6e38 \\
NGC1800-N &2, PA210\deg & 50$^{k}$&34.3 &1350  & 1.497e3	&6.16e-17	       &2.0 &	49   &16	&9.1e40	 \\
NGC2363-A &1, 5, 6&	44	&34     &593   & 1.533e2	&9.68e-17	       &0.76 &	15  &7.9	&1.2e40 \\	  
	B & 5  &	69	&38.4   &670   & 1.023e2	&3.66e-16	       &1.04 &	49  &5.7	&5.9e40 \\
NGC1569-C  &10,11   &	79	&8.07   &85    & 2.096e2	&1.45e-14	       &0.23 &	14   & 0.6	&1.4e39	 \\
	A &10,13,4,7 &	120	&105    &1120  & 3.960e3	&4.69e-16	       &1.7 &	250   & 5.5	&8.6e41	 \\
	B &11,6 &	85	&90.3   &960   & 4.451e3	&3.13e-16	       &1.6 &	120   & 6.6	&2.2e41	 \\
	D &10,6 &	54	&98.4   &1050  & 2.651e3	&4.66e-16	       &0.84 &	24   & 11	&6.9e40	 \\
	E &12, 6&	44	&76.6   &820   & 3.410e3	&2.48e-16	       &1.1 &	20   & 11	&2.3e40	 \\
	F &4, 13, 7&	64	&83.8   &890   & 3.035e3	&4.61e-16	       &2.2 &	89   & 8	&8.2e40	 \\
	G &4, 7, 11, 13&64	&99.5   &1060  & 2.736e3	&3.66e-16	       &1.2 &	49   & 9.7	&1.2e41	 \\
NGC2537-A &1, 3&	47	&4.8    &175   & 1.095e2	&8.42e-17	       &0.087 &	1.9    &2.2	&1.3e39	 \\
   ~~big  &    &	47	& 22.1  & 800  & 3.118e2 	& 2.96e-17	       &0.10 &	 2.3  &10	& 2.6e40 \\
NGC3077-D  &3,2   &	106	&17.7   &310   & 3.263e2	&3.29e-15	       &0.32 &	35   & 1.7	&4.5e40	 \\
	A & 5   &	55	&46.2   &810   & 1.298e3	&5.71e-16	       &1.4 &	43   & 8.6	&4.3e40	 \\
	J & 2,4   &	67	&53.8   &940   & 1.342e3	&2.91e-16	       &1.0 &	45   & 8.2	&1.1e41	 \\
	B & 5,2,3 &	40	&36.9   &640   & 6.256e2	&1.23e-15	       &0.93 &	15   & 9.4	&1.0e40	 \\
	G &2    &	51	&47.8   &830   & 9.076e2	&4.06e-16	       &0.61 &	16   & 10	&3.6e40	 \\
NGC4861-A  &1, 4   & 	50      &23.4   &850   & 5.024e2 	&3.9e-16               &2.7 &	66   & 10  &3.6e40  \\
        C &2, 4    & 	30      &26.4   &960   & 2.894e2	&3.8e-16               &1.3 &	12   & 19  &9.9e39  \\
	B &1    &	43	&15.7   &570   & 3.065e2	&7.8e-16	       &1.2 &	23   & 7.8	&1.0e40	 \\
NGC5253-E& PA60\deg &	35$^{k}$& 43.8  & 870  & 1.935e3	&2.37e-16	       &1.2 &	 14   &15   &1.3e40  \\
NGC4214-A  &1,2,3   &	45	&8.33   &150   & 1.350e2	&1.27e-14	       &0.13 &	2.8   & 2	&7.6e38	 \\
	B &1    &	28	&30.3   &530   & 6.698e2	&1.69e-15	       &1.1 &	8.3   & 11	&2.4e39	 \\
	C&1     &	100	&30.7   &540   & 2.597e2	&1.13e-15	       &0.41 &	41   & 3.2	&1.1e41	 \\
	D &4    &	46	&14.2   &250   & 6.597e1	&3.58e-15	       &0.14 &	3.0   & 3.2	&2.4e39	 \\
	F &2	&       44      & 37.4  &630   & 6.869e2	& 1.01e-15	       &0.61 &	12   & 8.4&1.3e40	 \\
NGC4449-A  &1,2,3   &	79	&43.1   &950   & 1.440e3	&5.82e-16	       &1.5 &	95   & 7.1	&1.8e41	 \\
	C & 1,2,3,11  &	30	&26.8   &470   & 2.163e3	&9.49e-16	       &3.4 &	31   & 9.2	&2.4e39	 \\
	D &9,11    &	33	&13.6   &240   & 7.295e2	&5.44e-16	       &0.49 &	5.3   & 4.3	&8.2e38	 \\
	G &9    &	25	&14.4   &250   & 1.642e3	&8.12e-16	       &1.7 &	11   & 5.9	&3.9e38	 \\
	H &7    &	51	& 43.9  & 770  & 1.045e3	& 4.92e-16	       &1.0 &	27   & 8.9	&3.1e40	 \\
	I &7    &	50	& 25.7 	& 450  & 1.067e3	& 1.25e-15	       &1.4 &	 35  & 5.3	&1.0e40	 \\
	E4&10    &	27	& 3.4   & 60   & 4.473e1	& 5.04e-16	       &0.012 &	 0.090  & 1.3	&2.8e37	 \\
	E2& 7   &	31	& 7.2   & 126  & 2.702e2	& 8.08e-16	       &0.090 &	 0.86  & 2.4 &1.9e38	 \\
	E3& 10   &	146	& 7.7   & 130  & 2.654e2	& 8.49e-16	       &0.14 &	 31  & 0.5	&2.1e40	 \\
	E1& 10   & 	40      & 4.5   & 79   & 2.158e2	& 3.07e-16	       &0.060 &	 0.96  & 1.2	&1.6e38	 \\
M82-N	  &2, 4&	156	&96.2   &1679  & 3.681e4	&4.28e-16	       &55 &	13400   &6.3	&4.2e42	 \\
	S &2   &	135	&126.7  &2210  & 1.999e4	&6.32e-16	       &29 &	5300   & 9.6	&4.8e42	 \\
\tableline
\end{tabular}
\tablenotetext{a}{(col. 1) Name given to each expanding complex of warm, ionized filaments
(see \fig~\ref{fig:images}).}
\tablenotetext{b}{(col. 2) Slits along which the complex was kinematically detected. }
\tablenotetext{c}{(col. 3) Maximum expansion velocity along our sightline. }
\tablenotetext{d}{(cols. 4, 5) Projected distance of shell from star-forming region. }
\tablenotetext{e}{(col. 6) Area of aperture in Figure~2. }
\tablenotetext{f}{(col. 7) Surface brightness  (\Ha + [NII]) of emission-line filaments within the polygonal
apertures shown in \fig~\ref{fig:images}.  Units are ergs/s/cm$^2$/asec$^2$. }
\tablenotetext{g}{ (col. 8) Upper limit on mass of warm ionized gas in the expanding complex, where 
$M = 14/11~  n_{rms} m_{\rm H} V \epsilon^{1/2}$.
Warm ionized clouds are assumed to fill a fraction
$\epsilon = 0.1$ of the bubble volume, $V$; and the root mean square electron
density is derived from \Ha\ photometry (see \S~\ref{sec:smass}). }
\tablenotetext{h}{(col. 9) Estimate of kinetic energy in expanding shell, $KE = 0.5 M v^2$.}
\tablenotetext{i}{(col. 10) Dynamical age of shell, $\tau = 0.6 R / v$. }
\tablenotetext{j}{(col. 11) Dynamical estimate of kinetic energy injection rate (see \S~\ref{sec:power}).  
The inferred power scales linearly with the average ambient hydrogen density
$\bar{n}(1)$. }

\tablenotetext{k}{Measurement from Marlowe \et (1995).}

\end{table}
\newpage
\begin{table}[h]

\caption{Global Star Formation Rates}
\begin{tabular}{llllll}
\hline
\hline
Galaxy	& F(\Ha+[NII])~\tablenotemark{a}
	& c(\Hb)~\tablenotemark{b}	
	& [NII]/\Ha~\tablenotemark{c}
        & Q~\tablenotemark{d}
 	& \.{M}~\tablenotemark{e} 	\\
	&(ergs/s/cm$^2$)	& (mag)	&		& (s$^{-1}$)	& \msunyr\	\\
\hline
VII~Zw~403& $5.94 \pm 0.2e-13 $	& 0.1	& 0.027		& 7.68e50	& 0.0024 \\
Sex~A-EW& $> 2.44 \pm 0.55e-12$~\tablenotemark{f}  
                    		& 0.021	& 0.021		& 6.58e50~\tablenotemark{g} & 0.0020 \\
I~Zw~18	& $4.44 \pm 0.1e-13 $	& 0	& 0.004		& 3.87e51	& 0.012 \\
II~Zw~40& $2.23 \pm 0.90e-12 $	& 1.1	& 0.026		& 1.05e53	& 0.33 \\
\n3738	& $1.40 \pm 0.05e-12 $	& 0.5	& 0.063		&3.97e51	& 0.013 \\
\n1800	& $6.30\pm 2.0e-13 $	& 1.0	& 0.150		& 1.45e52	& 0.046	\\	
\n2363	& $5.04 \pm 0.2e-12 $	& 0.1	& 0.026		& 6.49e51	& 0.021 \\
\n1569	& $2.31\pm 1.1e-11 $	& 0.83	& 0.025		& 3.41e52	& 0.11 \\
\n2537	& $7.98 \pm 0.83e-14 $	& 0.3	& 0.160		& 5.37e50	& 0.0017 \\
\n3077	& $1.74 \pm 0.08e-11 $	& 0.8	& 0.256		& 5.36e52	& 0.170 \\
\n4861	& $2.75 \pm 0.4e-12$	& 0.07	&0.019		&1.48e52	& 0.047 \\
\n5253	& $1.53e-11 $		& 0.4	& 0.077		&3.86e52	& 0.12 \\
\n4214	& $1.47 \pm 0.16e-11 $	&0.14	&0.062		&1.94e52	& 0.061 \\
\n4449	& $2.42 \pm 1.7e-11 $	& 0.27	&0.127		&3.68e52	& 0.12 \\
\hline
\end{tabular}

\tablenotetext{a}{(col. 2) Total \Ha + [NII] line flux corrected for atmospheric
extinction.  The narrowband images were obtained at the Steward Observatory Bok
telescope, and the data reduction and calibration are described in
Martin \& Kennicutt (1995).}
\tablenotetext{b}{(col. 3) Logarithmic extinction at \Hb (Martin 1997). }
\tablenotetext{c}{(col. 4) Intensity ratio of [NII]$\lambda6584$ emission to \Ha emission (Martin 1997). }
\tablenotetext{d}{(col. 5) Luminosity of hydrogen ionizing photons assuming Case~B recombination  at
$T_e = 10^4$~K. }
\tablenotetext{e}{(col. 6) Formation rate of 1 to 100\msun\ stars. }

\tablenotetext{f}{Includes only the two brightest HII regions.}
\tablenotetext{g}{Derived from the integrated \Ha luminosity
(Hunter \& Plummer 1996),
which is $9 \times 10^{38}$\ergsec after correcting for
foreground extinction.}

\label{tab:star}
\end{table}

\newpage
\begin{table}[h]
\scriptsize

\caption{Galactic Environment and Global Properties of the Feedback}
\begin{tabular}{lllllllllll}
\tableline
\tableline
Galaxy 	&  $M_{dyn}$~\tablenotemark{a}
      	& $M_{HI}$~\tablenotemark{b}
	& $L_B$~\tablenotemark{c}    
	& Max Disk~\tablenotemark{d}	
	& $L_x$~\tablenotemark{e}	
	& $M_x f^{1/2}$~\tablenotemark{f}	
	& $M_w \sqrt{\epsilon/0.1}$~\tablenotemark{g}	
	& $dM/dt \sqrt{\epsilon/0.1}$~\tablenotemark{h} 
	& $KE/E_B$~\tablenotemark{i}
 	& $\tau_{ej} / \tau_*$~\tablenotemark{j}	\\
		&  (\msun)   	
		& (\msun)	
		& (\lsun)  
		& $(M/L)_{*}$ 
		& (ergs/s)	
		& (\msun)	
		& (\msun) 
		& (\msun / yr) 
		&  & \\
\tableline    		 			    	 
VII~Zw~403 & \nodata	&	4.9e7 	& 6.0e6	  & \nodata & $< 2e38$   	&\nodata& \nodata	& 0     &\nodata& \nodata\\
Sex~A   & 3.2e8    	&	1.1e8	& 7.0e6	  & 24      & \nodata   	&\nodata& 2.8e4	(0.19)	& 0.034 &0.0003 & 22\\ 
I~Zw~18  & 5.0e8    	&	7e7	& 7.0e6	  & 57      & $< 1e39$ 		&\nodata & 4.7e5 (0.33) 	& 0.028  & 0.001& 33 \\
II~Zw~40 & 2.5e9\tablenotemark{k}&4.5e8	& 1.3e7   & 144     & \nodata 	& \nodata& \nodata		& 0	& \nodata   	& \nodata\\
NGC~3738& \nodata   	&	1.6e8	& 5.1e7	  & \nodata & \nodata   	&\nodata& 2.1e5 (0.86)	& 0.0020& \nodata & \nodata \\
NGC~1800& 2.1e9	    	&	1.6e8	& 9.4e7	  & 20      & \nodata		& \nodata&2.7e6 (0.38) 		& 0.14 	& 0.006	& 2 \\
NGC~2363& 3.6e9    	&	9.9e8 	& 1.0e8	  & 22      & \nodata   	&\nodata& 1.8e6 (0.65)	& 0.23 	& 0.006	& 11 \\
NGC~1569& 2.2e8    	&	8.4e7	& 1.6e8   & 1.08    & 1.9e38		& 1.4e6	& 8.8e6 (0.10)	& 0.93 	& 0.3	& 0.7\\
NGC~2537&  \nodata	&	2.8e8	& 1.8e8	  & \nodata & \nodata   	& \nodata& 9.4e4 (0.55)	& \nodata & \nodata& \nodata \\
NGC~3077& 5.9e9    	&	1.2e9	& 2.3e9	  & 3.4     & \nodata   	&\nodata& 4.3e6 (0.09)	& 0.43  & 0.001 & 6 \\
NGC~4861& 7.1e9    	&	5.4e8	& 7.3e7	  & 87      & \nodata   	&\nodata& 5.2e6 (0.23)	& 0.27  & 0.004 & 1\\
NGC~5253& $> 4.3e8$	&	1.4e8	& 2.3e8	  & $\sim 1$&$< 6.5e38$   &$< 1.5e5$ &1.2e6 (0.59)	& 0.08  & 0.003 & 12 - 19\\
NGC~4214& 3.3e9    	&	1.8e9	& 2.3e8   & 3.4     & \nodata   	&\nodata& 2.4e6 (0.12)	& 0.22  & 0.0003& 43 \\
NGC~4449& 2.6e10    	&	2.2e9	& 2.8e8	  & 80      & 6.5e38   	& 6.1e6	& 9.93e6 (0.08)		& 1.7   & 0.0008	& 2 \\
M~82    & 8.6e9    	&	8.8e8	& 7.7e8	  & 10      & 3e40     	& 1.47e8 & 8.5e7 (0.13)		& 24    & 0.04	& 0.5  \\ 
\tableline	
\end{tabular}

\tablenotetext{a}{(col 2) Keplerian mass within $R_{HI}$. }
\tablenotetext{b}{(col 3)  Atomic hydrogen mass. References listed in Table~2. }
\tablenotetext{c}{(col 4) Galactic B-band luminosity in solar units. }
\tablenotetext{d}{(col 5) Implied stellar mass to blue light ratio in absence of
dark matter, where  $(M/L)_{*} \equiv (M_{dyn} - 1.4 M_{HI}) / L_B$ in solar units.}

\tablenotetext{e}{(col 6) Unabsorbed soft X-ray luminosity.  Point sources are excluded
if they are resolved.  Bandpasses and references as follows:
\n1569: 0.5 -- 2.0 keV (Della Ceca \et 1996);
IZw18:  0.1 -- 2.2 keV (Martin 1996);
M82:    0.2 -- 3.5 keV (Strickland \et 1997);
VIIZw403: 0.1 -- 2.4 keV (Papaderos \et 1994);
\n4449: 0.5 -- 2.0 keV (Della Ceca \et 1997); and
\n5253: 0.1 -- 2.4 keV (Martin \& Kennicutt 1995). }

\tablenotetext{f}{(col 7) Mass of hot, diffuse ISM defined as $M_x = (7/6) n_e m_H V$.
The electron density was estimated from the spectral normalization and volume given
in Della Ceca \et 1996 (\n1569), Della Ceca \et 1997 (\n4449), 
Martin \& Kennicutt 1995 (\n5253),
and Strickland \et 1997 (M82).
All values are scaled to the distance given in Table~1 and a He:H ratio of 1:10 by number. }

\tablenotetext{g}{(col 8) Upper limit on the total mass of warm ionized gas in large shells.
i.e. The shell masses in column~8 of Table~3 have been added together assuming
a filling factor  $\sqrt{\epsilon/0.1}$, but $\epsilon$ may be considerably smaller.
The fractional uncertainty, in parentheses, reflects the uncertainty in the volume estimate  
(see \S~\ref{sec:smass}).}

\tablenotetext{h}{(col 9) Estimated mass ejection rate from the galactic  disk,
 $dM/dt \equiv (M_w + M_x)  / \tau_{dyn}$.  Note that the contribution from
hot gas is only included for M82, \n1569, and \n4449. }

\tablenotetext{i}{(col 10) Upper limit on ratio of kinetic energy in large-scale ionized hydrogen
shells to the gravitational binding energy of the atomic gas.
The binding energy estimate, $E_B \sim (1.4 M_{HI}) v_{c}^2$, is  a lower
limit; and the kinetic energy scales as $\sqrt{\epsilon/0.1}$ where
$\epsilon \sles 0.1$. }

\tablenotetext{j}{(col 11) Fraction of an orbital period to eject the entire ISM at the
rate given in col.~9, where $\tau_{ej} =  1.4 M_{HI} / dM/dt$ and
$\tau_* = 2 \pi R(v_{c}) / v_{c}$. }

\tablenotetext{k}{Mass of northern cloud.  Southern cloud adds an additional
$6 \times 10^9$\msun.}

\normalsize
\label{tab:mdot}
\end{table}



\clearpage\newpage

%

%
%
%
%

\newpage

\begin{figure}[p]
\caption[\Ha echellograms]{\Ha echellograms.  The ticks mark increments of 
46 \kms in the horizontal direction and 60\arcsec\ in the vertical direction.
Wavelength increases to the right.  The name of the galaxy and label 
identifying the position in Table~\ref{tab:bubbles}
are printed at the top of each slit.}
\label{fig:edata}
\end{figure}

\begin{figure}
\caption[Location of Kinematic Shells]
{Positions of expanding shells  relative to the \Ha
morphology. North is up, and east is to the left.  Ellipses
mark the locations of the Doppler ellipses identified in the
echellograms.  The  axis of the ellipse colinear with the position
angle of the slit represents the spatial dimension of the Doppler
ellipse, and the second axis (usually the minor axis) is scaled such
that 1 pixel is  approximately  0.73 \kms of velocity splitting.
The galaxies are:
(a) \n3077, (b) \n4214, (c) \n4861, (d) M82, (e) \n3738, (f) \n2537,
(g) \n2363, (h) \n1800, (i) Sex~A, (j) \n5253, (k) \n1569, and
(h) \n4449.
}
\label{fig:images}
\end{figure}

\begin{figure}
\caption[Distribution of shell sizes and radial velocities.]
{Measured radii and radial velocities of kinematic shells.
For comparison, the
open symbols show the kinematic shells found by Marlowe \et (1995).
}
\label{fig:rv}
\end{figure}
\begin{figure}
\caption[Bubble ages and power requirements.]
{Estimated bubble ages and  power requirements.
The dashed lines are isochrones  for adiabatic, wind-driven bubbles
with radiating shells.  
A shell evolves parallel to a  dotted  line if the wind is steady.
The model assumes the ambient medium is homogeneous.
The curves of constant mechanical power are normalized to an ambient hydrogen
number density of  $n = 1$\cm3, so the units for the implied power
are  $(n / 1$\cm3) \ergsec.  The symbol type identifies the
galaxy.
 The shell will be moving supersonically, $v \ge 20$\kms,
 when it reaches one scale-height, $h = 500$~pc, if it
 lies above the thick dashed line (green).
}
%
\label{fig:weaver}
\end{figure}
\begin{figure}
\caption[Constraints on the average ambient H density.]
{Comparison of kinematic energy estimates for ionized shells.
For the y-coordinate, the shell mass was derived from the bubble's
volume and an ambient density $n = 0.1$\cm3.  For the x-coordinate,
the total mass of ionized hydrogen in the complex was computed from
areal \Ha photometry and a filling factor for warm, ionized clouds
of $\epsilon = 0.1$.
}
\label{fig:ke}
\end{figure}
\begin{figure}
\caption{
Power requirements of the expanding shells versus the current
mechanical power production from massive stars in the galaxy.
The y-axis is scaled to an average ambient density of 0.1\cm3.
}
\label{fig:stars}
\end{figure}
%
%

\begin{figure}
\caption{
Rotation curve of \n1569.  Open and starred symbols are based on the
HI maps of Reakes (1980) and Israel \& van Driel (1990), respectively.
The position - velocity diagrams were extracted along ${\rm PA} = 117\deg$
and deprojected assuming an inclination of $\iota = 63\deg$.
Triangles and squares denote points to the southwest and northeast of
the center of the galaxy, respectively.   The dotted and dashed lines are the circular
velocity curves of a Burkert halo model with core radius $r_0 = 1.4$~kpc and 
532~pc, respectively.  Increasing
the central density by factors of 3.4 and 4.5 from Burkert's empirical
$\rho_0(r_0)$ relation yields the higher circular velocities 
(solid line).}
\label{fig:rotate}
\end{figure}
\begin{figure}
\caption
{
Escape velocity versus height above the center
of the disk of \n1569. Solid points show the radius
and projected expansion speed of the supershells
for comparison. Three models for the potential are
shown:  (1) {\em Dashed line} -- exponential disk
with 500~pc scalelength and central surface density
240\msun~pc$^{-2}$,
(2) {\em Dotted line} -- sum of Burkert $r_0 = 1.4$~kpc 
halo and exponential disk with $\alpha^{-1} = 500$~pc,
$\sigma_0 = 155$\msun~pc$^{-2}$, and (3) {\em Solid line}
-- equal disk and halo mass within a radius of 1~kpc,
same scalelengths as (2).
Halo models were truncated at a radius of 160~kpc.
}
\label{fig:potent}
\end{figure}
\begin{figure}
\caption{
Rotation curves, halo escape velocity estimates,
and shell expansion velocities. Open symbols show HI position -- velocity data
described in Table~\ref{tab:sam}. Thin lines compare the rotation curves of 
Burkert halo models, and bold lines illustrate the halo escape velocity as a
function of galactocentric radius.
Only the dotted lines denote single parameter fits (Burkert's $\rho_0(r_0)$ 
relation).  For \n3077 and \n4214, the escape velocity above a disk model 
is also shown, dashed lines. 
Solid circles are the supershells.
}
\label{fig:vall}
\end{figure}
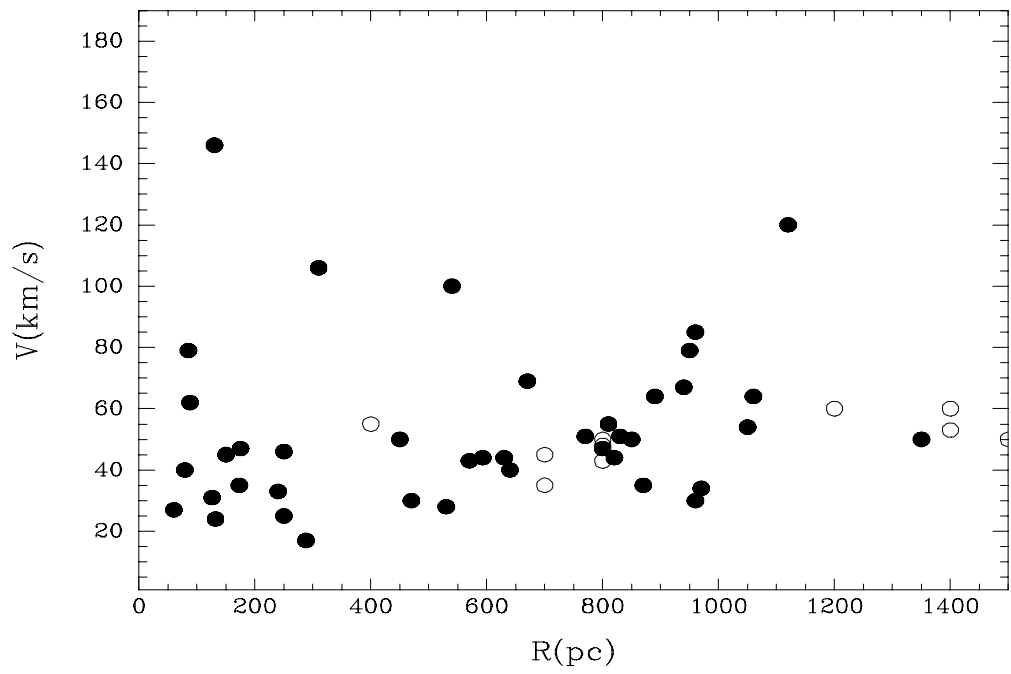
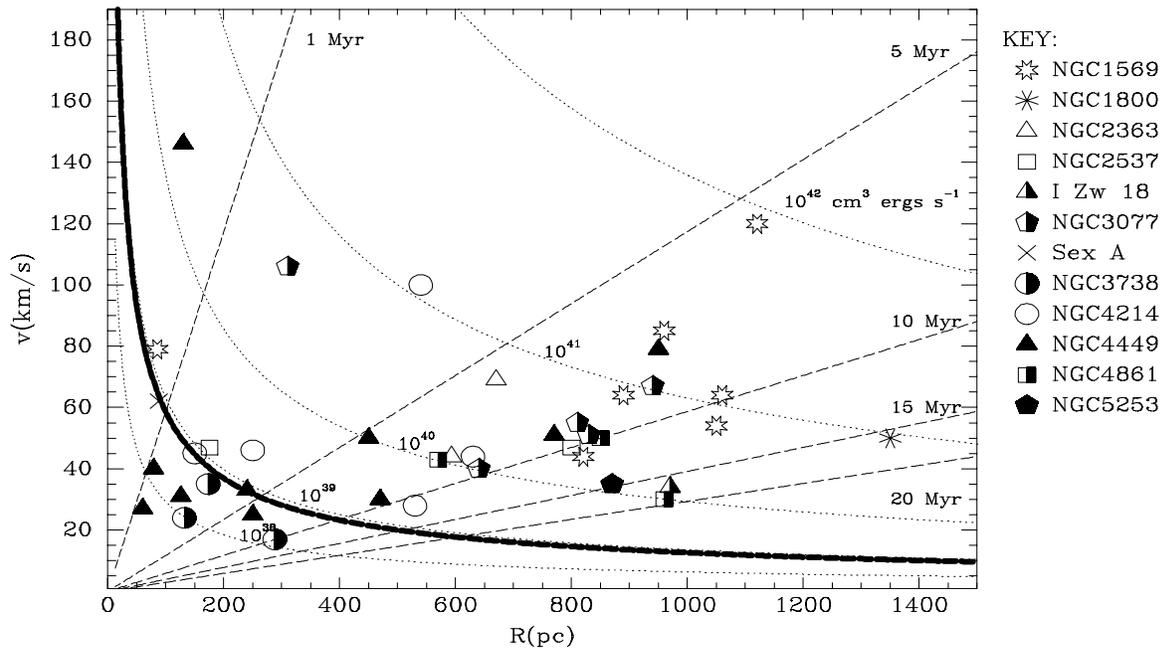
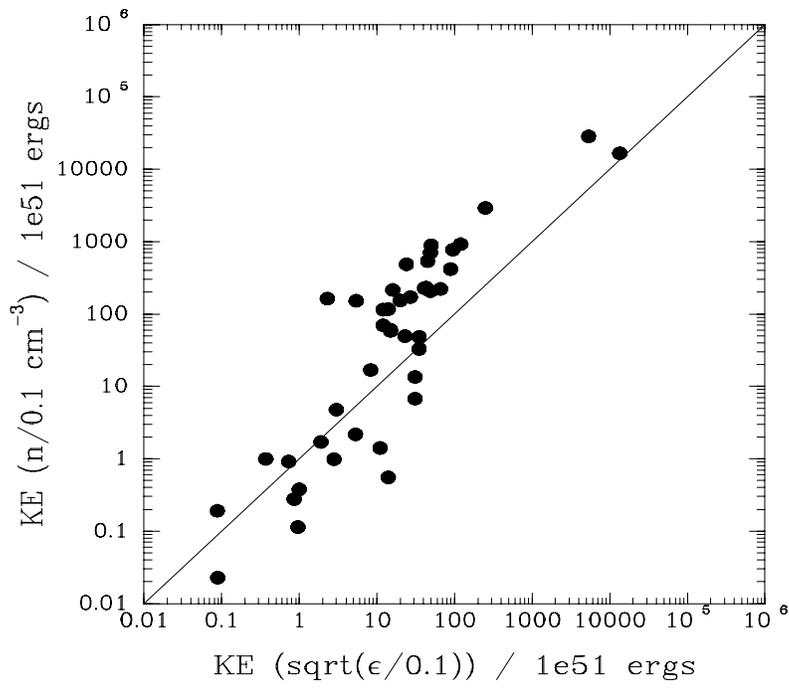
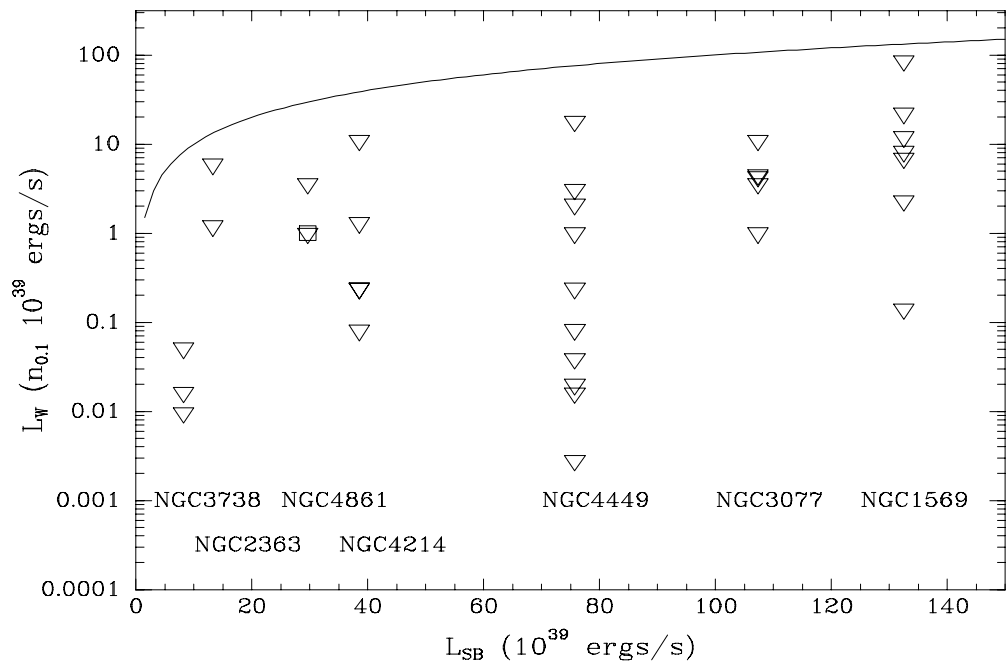
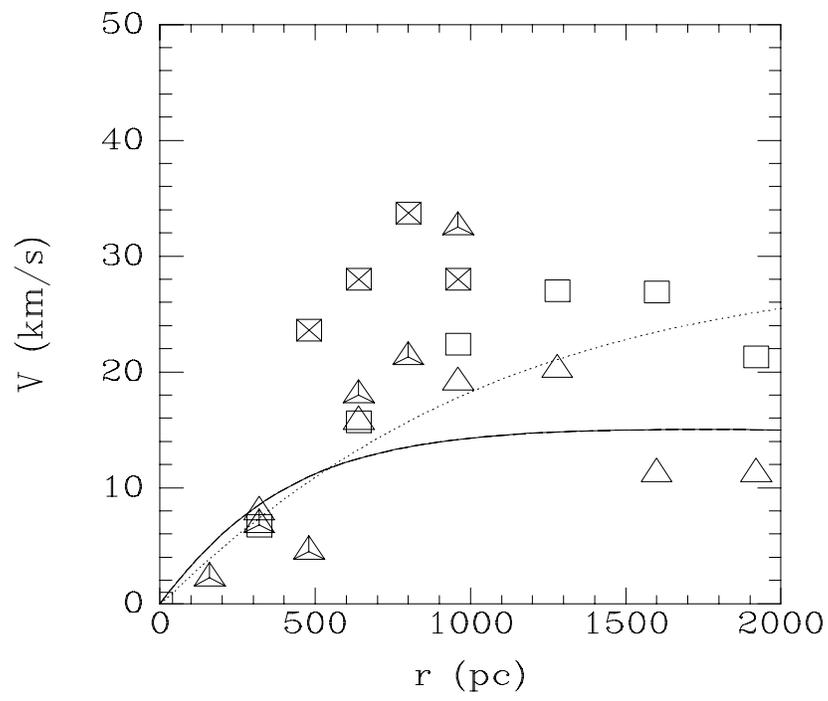
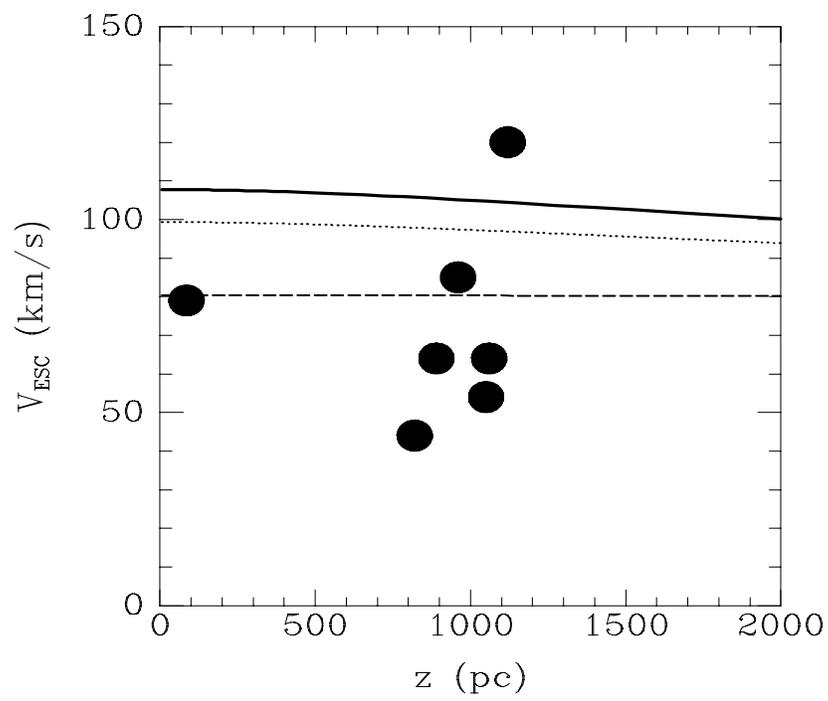
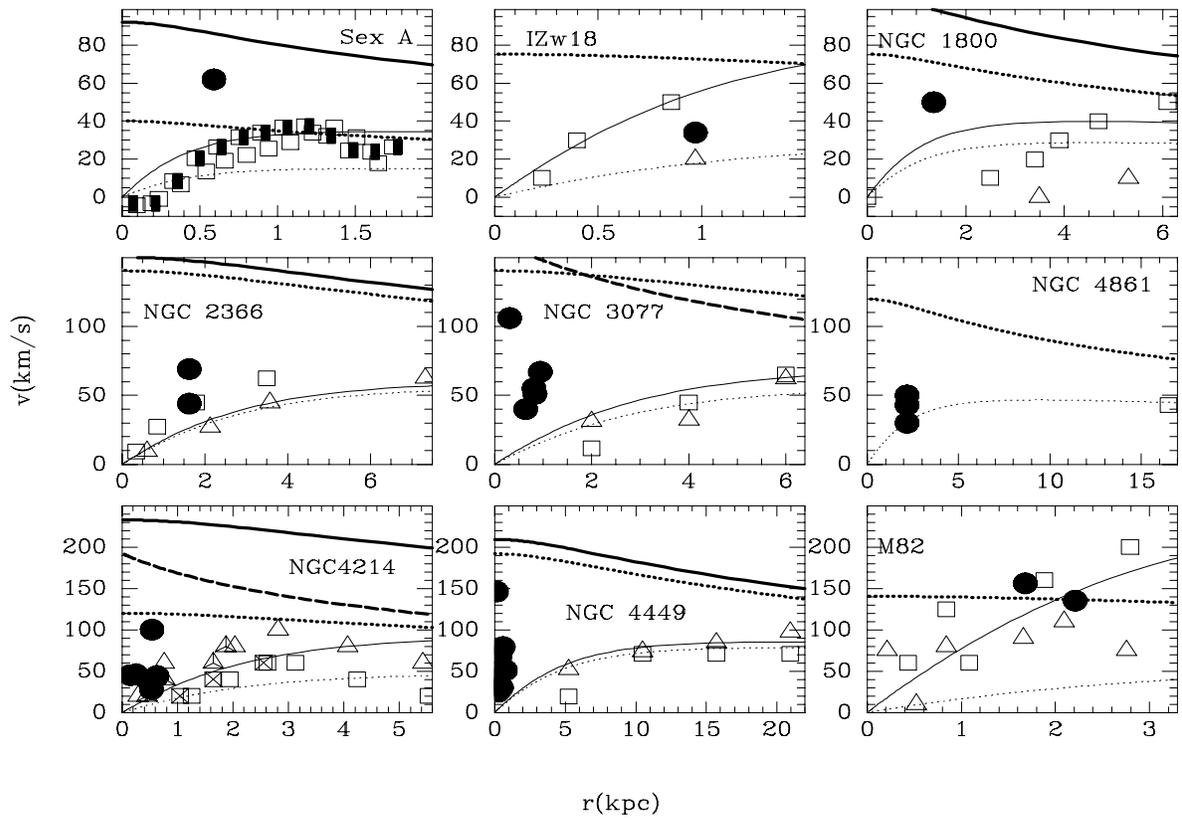
\end{document}